\documentclass[showpacs,amssymb,floatfix,pre,aps,notitlepage]{revtex4-1}

\usepackage{amssymb,amsfonts,bm}
\usepackage{graphics,graphicx}

\usepackage{colordvi,amsmath,epsfig,color}

\usepackage[activeacute,english]{babel}
\newcommand{\be}{\begin{equation}}
\newcommand{\ee}{\end{equation}}
\begin{document}

\title{Linear hydrodynamics for driven granular gases}

\author{Mar\'ia Isabel Garc\'ia de Soria}
\author{Pablo Maynar}
\affiliation{F\'{\i}sica Te\'{o}rica, Universidad de Sevilla,
Apartado de Correos 1065, E-41080, Sevilla, Spain}
\affiliation{F\'{\i}sica Te\'{o}rica, Universidad de Sevilla,
Apartado de Correos 1065, E-41080, Sevilla, Spain}
\author{Emmanuel Trizac}
\affiliation{Laboratoire de Physique Th\'eorique et Mod\`eles Statistiques, 
UMR CNRS 8626, Universit\'e Paris-Sud, 91405 Orsay, France}

\begin{abstract}
We study the dynamics of a granular gas heated by the stochastic thermostat. 
From a Boltzmann description, we derive the hydrodynamic equations for small 
perturbations around the stationary state that is reached in the long time 
limit. Transport coefficients are identified as Green-Kubo formulas 
obtaining explicit expressions as a function of the inelasticity and the 
spacial dimension. 
\end{abstract}
\maketitle

\section{Introduction}

Granular assemblies --made up of macroscopic solid bodies undergoing dissipative interactions-- 
have been shown \cite{c90, h83} to frequently exhibit flows 
similar to those of normal fluids, and 
for practical purposes are often described by phenomenological 
hydrodynamic equations, i.e. equations for the density, flow velocity and 
energy density. This is so even if energy is not a conserved variable (the 
condition is that the energy mode be a slow variable compared to the rest of 
excitations). When the dynamics of the grains 
can be partitioned into sequences of two-body collisions, there is support both 
from experiments and computer simulations for the usefulness of a Kinetic 
Theory description. This occurs, schematically, when there are no clusters nor 
jamming and the system remains fluidized \cite{l05}. In the low density limit, 
the relevant dynamics is encoded in the one-particle distribution function, 
which obeys the inelastic Boltzmann equation 
\cite{gs95, bds97}. This is the starting point for many of the formal 
derivations of 
hydrodynamic equations by applying similar tools and ideas as those used in the context 
of ordinary fluids \cite{resibois}. In 
the free-cooling case the study of the existence and applicability of a 
hydrodynamic regime is rather complete for the inelastic hard sphere (IHS) 
model. The Navier-Stokes equations have been derived by the 
Chapman-Enskog expansion \cite{bdks98} and also via the linearized Boltzmann 
equation \cite{db03, bdr03, bd05}, yielding equivalent Green-Kubo formulas for 
the transport coefficients \cite{db02}. 
The problem for arbitrary densities has also been  tackled in 
\cite{dab08, adb08} applying linear response methods. Although the successful 
of the 
Navier-Stokes equations is remarkable, granular systems often require to go beyond 
this level of description. For these cases and close to a stationary state, a 
modification of the Chapman-Enskog expansion has been carried out taking 
into account rheological effects \cite{l06, g06}.  

Some modifications of the IHS model have been introduced for the study of other 
situations, mainly when a stationary state is reached. For example, when in 
a vibrated system the stationary state is quasi-homogeneous, or 
when the grains are immersed in an interstitial medium that acts as a 
thermostat \cite{peu02, lvu09, pggsv12,gtsh12}, the system can be effectively 
modelled as driven by some random energy source. 
The energy injection can be performed by applying a random force to each 
particle. Depending on the stochastic properties of this force, different kinds 
of thermostats are obtained \cite{clh00}, one of the most used being the 
so-called stochastic thermostat, which consists of a white noise force acting 
on each grain \cite{wm96, vne98, vnetp99, PTvNE01, ms00, gm02, etb06, vpbtvw06, vpbvwt06, 
gmt09, mgt09, vaz11, plmv99v, Bada2012}. This model has been less studied 
that its unforced version (free-cooling 
case). 
To our knowledge, the only derivations of the 
hydrodynamic equations are the one made in \cite{gm02} which consider 
some variants of the present stochastic thermostat in which the heating may
depend on the local temperature. The objective of this work is to go further 
in this direction and to derive the hydrodynamic equations for the actual 
homogeneous stochastic thermostat model in the low density limit. 
It is arguably the most commonly employed model in the simulations. At 
the Boltzmann equation level, we will consider states that are close to the homogeneous 
stationary regimes, thereby obtaining linear equations for the deviations of the 
hydrodynamic fields around their homogeneous counterparts, with explicit 
Green-Kubo formulas for the transport coefficients. Let us note that, very 
recently, a study in the same lines has been done in \cite{Bada2012} by 
a different method, applying the Chapmann-Enskog expansion to the Enskog 
equation. 

The plan of the paper is as follows. The model is first defined in section \ref{section2}.
We summarize the main properties of 
the most general hydrodynamic state through which the stationary regime is 
reached in the long time limit. This state was analyzed in detail in 
\cite{gmt12} and, as will be shown, plays an essential role in the hydrodynamic 
description for the present model. In section \ref{section3}, the linearized Boltzmann 
equation is written and the relevant modes for the hydrodynamic description are 
identified. These properties are subsequently exploited in section \ref{section4} to derive 
the linearized hydrodynamic equations. Finally, in section \ref{section5}, we 
present a short summary of the results obtained in the paper while  
details of the calculations are given in seven appendixes, at the end of the 
text.

\section{The model}
\label{section2}

Let us consider a dilute gas of $N$ smooth inelastic hard spheres 
($d=3$) or disks ($d=2$) of mass $m$ and diameter $\sigma$. These bodies collide 
inelastically with a 
coefficient of normal restitution $\alpha$, independent of the relative 
velocity. If at time $t$ there is a binary encounter between particles 
$i$ and $j$, with velocities $\mathbf{V}_i(t)$ and $\mathbf{V}_j(t)$ 
respectively, the postcollisional velocities $\mathbf{V}_i'(t)$ and 
$\mathbf{V}_j'(t)$ are
\begin{eqnarray}\label{collisionRule}
\mathbf{V}_i'&=&\mathbf{V}_i-\frac{1+\alpha}{2}
(\hat{\boldsymbol{\sigma}}\cdot\mathbf{V}_{ij})\hat{\boldsymbol{\sigma}},
\nonumber\\
\mathbf{V}_j'&=&\mathbf{V}_j+\frac{1+\alpha}{2}
(\hat{\boldsymbol{\sigma}}\cdot\mathbf{V}_{ij})\hat{\boldsymbol{\sigma}},
\end{eqnarray}
where $\mathbf{V}_{ij}\equiv\mathbf{V}_i-\mathbf{V}_j$ is the relative
velocity and $\hat{\boldsymbol{\sigma}}$ is the unit vector pointing from the
center of particle $j$ to the center of particle $i$ at contact. Between 
collisions, the system is heated uniformly by adding a random velocity to the 
velocity of each particle independently with certain frequency and 
with a given probability distribution. Let us define the jump distribution, 
$P_{\Delta t}(\Delta\mathbf{v})$, as the probability that a particle experiences 
a jump $\Delta\mathbf{v}$ in the time interval $\Delta t$, that will be assumed 
to fulfill 
\be
\lim_{\Delta t\to 0}\int d\mathbf{y}y_jP_{\Delta t}(\mathbf{y})=0, \qquad
\lim_{\Delta t\to 0}\frac{1}{\Delta t}
\int d\mathbf{y}y_j^2P_{\Delta t}(\mathbf{y})=\xi_0^2, \qquad j=1,\dots,d, 
\ee
where $\xi_0^2$ is the strength of the noise. If the variance of this 
distribution is small compared to the velocity scale in which 
the one-particle distribution, $f(\mathbf{r},\mathbf{v},t)$, varies, the 
evolution equation is, in the low density limit, the 
Boltzmann-Fokker-Planck equation \cite{vne98, vk92}
\be\label{ec.b}
\left[\frac{\partial}{\partial t}
+\mathbf{v}_1\cdot\frac{\partial}{\partial{\mathbf{r}_1}}\right]f(x_1,t)=
\sigma^{d-1}\int dx_2\delta(\mathbf{r}_{12})\bar{T}_0(\mathbf{v}_1,\mathbf{v}_2)
f(x_1,t)f(x_2,t)
+\frac{\xi_0^2}{2}\frac{\partial^2}{\partial\mathbf{v}_1^2}f(x_1,t),
\ee
where we have introduced the field variable $x\equiv\{\mathbf{r},\mathbf{v}\}$ 
and the binary collision operator $\bar{T}_0$ 
\be
\bar{T}_0(\mathbf{v}_1,\mathbf{v}_2)=\int d\hat{\boldsymbol{\sigma}}
\Theta(\mathbf{v}_{12}\cdot\hat{\boldsymbol{\sigma}})
(\mathbf{v}_{12}\cdot\hat{\boldsymbol{\sigma}})(\alpha^{-2}b_{\sigma}^{-1}-1).
\ee
Here the operator $b_{\sigma}^{-1}$ replaces the 
velocities $\mathbf{v}_1$ and $\mathbf{v}_2$ by the precollisional ones 
$\mathbf{v}_1^*$ and $\mathbf{v}_2^*$ given by 
\begin{eqnarray}
\mathbf{v}_1^*&=&\mathbf{v}_1-\frac{1+\alpha}{2\alpha}
(\hat{\boldsymbol{\sigma}}\cdot\mathbf{v}_{12})\hat{\boldsymbol{\sigma}},
\nonumber\\
\mathbf{v}_2^*&=&\mathbf{v}_2+\frac{1+\alpha}{2\alpha}
(\hat{\boldsymbol{\sigma}}\cdot\mathbf{v}_{12})\hat{\boldsymbol{\sigma}}.
\end{eqnarray}
Under the conditions noted above, the evolution equation 
does not depend on the details of the distribution $P_{\Delta t}$, but only on 
its second moment, through the coefficient $\xi_0^2$. 

It is convenient to introduce the hydrodynamic fields in the standard 
kinetic theory fashion, as the first velocity moments of the one-particle 
distribution function
\begin{eqnarray}
n(\mathbf{r},t)&=&\int d\mathbf{v}f(\mathbf{r},\mathbf{v},t), \\
n(\mathbf{r},t)\mathbf{u}(\mathbf{r},t)
&=&\int d\mathbf{v}\mathbf{v}f(\mathbf{r},\mathbf{v},t), \\
\frac{d}{2}n(\mathbf{r},t)T(\mathbf{r},t)
&=&\int d\mathbf{v}\frac{m}{2}V^2f(\mathbf{r},\mathbf{v},t),
\end{eqnarray}
where $\mathbf{V}=\mathbf{v}-\mathbf{u}$ is the velocity of the particle 
relative to the local velocity flow, $\mathbf{u}$. Let us also introduce the 
local thermal velocity through
\be
v(\mathbf{r},t)=\left[\frac{2T(\mathbf{r},t)}{m}\right]^{1/2}. 
\ee
We will see that, in terms 
of the homogeneous hydrodynamic fields, we can specify some relevant states at 
the Boltzmann equation level. It is known numerically 
that for a wide class of initial conditions, the system 
reaches a homogeneous stationary state \cite{vnetp99}. Assuming that total momentum is zero, 
i.e. $\int d\mathbf{v}\mathbf{v}f(\mathbf{v},0)=\mathbf{0}$, the state is 
characterized by an isotropic stationary distribution, $f_s(v)$, which was 
studied in detail in \cite{vne98}. There, the distribution was written as
\be\label{fs}
f_s(v)=\frac{n}{v_s^d}\chi_s(c), \qquad 
\mathbf{c}=\frac{\mathbf{v}}{v_s}, 
\ee
where $n$ is the total density and $v_s$ is the stationary thermal velocity. 
We have also 
introduced the scaled distribution function, $\chi_s$, which is independent 
of the strength of the noise $\xi_0^2$, i.e. all the dependence of the 
distribution on $\xi_0^2$ is written in terms of the temperature. 
As this distribution is quite close to a Maxwellian, an expansion 
in terms of Sonine polynomials \cite{resibois} does make sense. In the 
so-called first Sonine approximation the function reads \cite{vne98}
\be
\chi_s(c)\approx\chi_M(c)[1+a_2^sS_2(c^2)], 
\ee
where $\chi_M$ is the Maxwellian distribution with unit temperature, 
$S_2(c^2)=\frac{d(d+2)}{8}-\frac{d+2}{2}c^2+\frac{1}{2}c^4$ is the second 
Sonine polynomial, and $a_2^s$ is the kurtosis of the distribution. Within this 
approximation and neglecting non linear contributions in $a_2^s$, the 
distribution function can be calculated as 
\cite{vne98}
\begin{equation}
a_2^s(\alpha)=\frac{16(1-\alpha)(1-2\alpha^2)}
{73+56d-24d\alpha-105\alpha+30(1-\alpha)\alpha^2},
\end{equation}
with a stationary temperature 
\begin{equation}\label{ts}
T_s=m\left[\frac{d\Gamma(d/2)\xi_0^2}
{2\pi^{\frac{d-1}{2}}(1-\alpha^2)n\sigma^{d-1}}\left(1-\frac{3}{16}a_2^s\right)
\right]^{2/3}. 
\end{equation}

It has recently been shown that, in a homogeneous situation, 
the initial condition is ``forgotten'' before the stationary state has been 
reached \cite{gmt12}. The system approaches the stationary state through 
a universal route in which all the time 
dependence of the distribution function, $f_H(\mathbf{v},t)$, goes through the 
instantaneous temperature, $T_H(t)$ (in the following we will denote the 
universal state with the subindex $H$). In contrast with the free cooling case, 
due to the parameter $\xi_0^2$, we can construct an additional quantity with 
dimensions of temperature apart from the instantaneous temperature, namely the 
stationary temperature given by eq. (\ref{ts}). Then, by dimensional analysis 
the distribution function has a two parameter scaling form 
\be\label{dps}
f_H(\mathbf{v},t)=\frac{n}{v_H(t)^d}\chi(c,\beta), \qquad 
\mathbf{c}=\frac{\mathbf{v}}{v_H(t)}, \qquad \beta=\frac{v_s}{v_H(t)}.  
\ee
Let us remark that we have redefined the variable $\mathbf{c}$ compared to the 
one introduced in eq. (\ref{fs}) and we have introduced the instantaneous 
thermal velocity, $v_H(t)$. In reference \cite{gmt12} it was 
shown that this state actually 
exists and the dynamics  is partitioned in a first rapid stage where initial 
conditions matter, and a subsequent universal relaxation towards stationarity, 
where only the distance to the steady state is relevant, through the 
dimensionless inverse typical velocity $\beta=v_s/v_H(t)$, i.e. 
\begin{equation}
f(\mathbf{v},t|f_0)\longrightarrow\frac{n}{v_H^d(t)}\chi(c,\beta)
\longrightarrow f_s(v). 
\end{equation}
Let us note that a similar two parameter scaling occurs in the uniform shear 
flow of granular gases \cite{as07, as12}. In this case, after a quick 
transient, the system 
forgets the initial condition and evolves to the stationary state through a 
normal state, where the role of $\beta$ is played by the dimensionless shear 
rate, $a^*= a [n\sigma^{d-1}v_0(t)]^{-1}$. As in the stationary state, numerical 
simulations show that the scaled distribution is close to a Maxwellian and it 
can be calculated in the first Sonine approximation
\be\label{sonApBeta}
\chi(c,\beta)\approx\chi_M(c)[1+a_2(\beta)S_2(c^2)], 
\ee
where, by definition, we have
\be
\int d\mathbf{c}\chi(c,\beta)=1, \qquad 
\int d\mathbf{c}\mathbf{c}\chi(c,\beta)=\mathbf{0}, \qquad
\int d\mathbf{c}c^2\chi(c,\beta)=\frac{d}{2}. 
\ee
Introducing the approximated distribution (\ref{sonApBeta}) into the Boltzmann 
equation, and neglecting the non-linear terms in $a_2(\beta)$, it is possible 
to identify the universal distribution, i.e. the universal $a_2(\beta)$, to be 
characterized by
\cite{gmt12}
\be
a_2(\beta)=a_2^s\left[1+\frac{1-\beta^3}{B-1}
\,_2F_1\left(-\frac{1}{3},1;\frac{4B-1}{3};\beta^3\right)\right],
\ee
valid for $0<\beta<1$ and 
\be
a_2(\beta)=-\frac{4Ba_2^s}{7\beta^3(1-1/\beta^3)^{\frac{4B}{3}}}
\,_2F_1\left(\frac{7}{3},1+\frac{4B}{3};\frac{10}{3};\frac{1}{\beta^3}\right),  
\ee
for $\beta>1$. Here we have introduced the hyper-geometric function, $\,_2F_1$, 
\cite{arnoldVasilii} and the coefficient
\be
B=\frac{73+8d(7-3\alpha)+15\alpha[2\alpha(1-\alpha)-7]}
{16(1-\alpha)(3+2d+2\alpha^2)
+a_2^s[85+d(30\alpha-62)+3\alpha(10\alpha(1-\alpha)-39)]}. 
\ee
As in the long time limit the distribution tends to the stationary state we have
\be
\lim_{\beta\to 1^+}a_2(\beta)=\lim_{\beta\to 1^-}a_2(\beta)=a_2^s. 
\ee 
We also have that the first derivative is continuous
\be
\lim_{\beta\to 1^+}\frac{d}{d\beta}a_2(\beta)
=\lim_{\beta\to 1^-}\frac{d}{d\beta}a_2(\beta), 
\ee 
an important property that will turn to be needed in the following sections. 

The evolution equation for the temperature (or equivalently for the thermal 
velocity, $v_H$) in the universal state can be calculated by inserting the 
scaling form (\ref{dps}) into the Boltzmann equation and taking the second 
velocity moment. When this is done, we obtain
\be\label{ec.vH}
\frac{dv_H(t)}{dt}=\ell^{-1}[\mu(1)\beta^3-\mu(\beta)]v_H(t)^2, 
\ee
where we have introduced the dimensionless coefficient
\be\label{def.mu}
\mu(\beta)=-\frac{1}{2d}\int d\mathbf{c}_1\int d\mathbf{c}_2
\chi(c_1,\beta)\chi(c_2,\beta)
T_0(\mathbf{c}_1,\mathbf{c}_2)(c_1^2+c_2^2). 
\ee
The operator $T_0$ is 
\be
T_0(\mathbf{c}_1,\mathbf{c}_2)=\int d\hat{\boldsymbol{\sigma}}
\Theta(\mathbf{c}_{12}\cdot\hat{\boldsymbol{\sigma}})
(\mathbf{c}_{12}\cdot\hat{\boldsymbol{\sigma}})(b_{\sigma}-1), 
\ee
where $b_{\sigma}$ replaces the velocities $\mathbf{v}_1$ and $\mathbf{v}_2$ by 
the postcollisional ones 
$\mathbf{v}_1'$ and $\mathbf{v}_2'$ given by 
\begin{eqnarray}
\mathbf{v}_1'&=&\mathbf{v}_1-\frac{1+\alpha}{2}
(\hat{\boldsymbol{\sigma}}\cdot\mathbf{v}_{12})\hat{\boldsymbol{\sigma}},
\nonumber\\
\mathbf{v}_2'&=&\mathbf{v}_2+\frac{1+\alpha}{2}
(\hat{\boldsymbol{\sigma}}\cdot\mathbf{v}_{12})\hat{\boldsymbol{\sigma}}.
\end{eqnarray}
By inserting the function $\chi$ in the first Sonine approximation, eq. 
(\ref{sonApBeta}), an approximate evolution equation for $v_H$ can be obtained. 
Nevertheless, we will be only interested on situations where we are close to 
the stationary state, $\beta=1$, and then equation (\ref{ec.vH}) can be 
linearized  obtaining
\be\label{evvH}
\frac{dv_H(t)}{dt}=-\gamma
\frac{v_s}{\ell}\left[v_H(t)-v_s\right], 
\ee
where we have introduced the dimensionless coefficient
\be\label{def.gamma}
\gamma=3\mu(1)-\left.\frac{d\mu(\beta)}{d\beta}\right|_{\beta=1}. 
\ee
The coefficient $\gamma$ is calculated in Appendix \ref{apendiceA} in the first 
Sonine 
approximation. Let us note that the difference of eq. (\ref{evvH}) with the 
equivalent one in \cite{vnetp99} is just the term 
$\left.\frac{d\mu(\beta)}{d\beta}\right|_{\beta=1}$, that, although small, 
is shown in the Appendix to be of the same order than $a_2$. The solution of 
equation 
(\ref{evvH}) is 
\be\label{ev.vH}
v_H(t)=v_s+\left[v_H(0)-v_s\right]e^{-\gamma\frac{v_s}{\ell}t}, 
\ee
that shows that the relaxation of $v_H$ to the stationary value is given in 
terms of the coefficient $\gamma$. 

The universal state that we have identified, that in the following will be 
referred to as the $\beta$-state, is the analogous, for heated 
granular gases, of the homogeneous cooling state for unforced systems \cite{bdks98}. It 
represents the most general homogeneous hydrodynamic state and, as will be seen 
in the next sections, its existence is crucial in the study of the relevance 
of a hydrodynamic description for granular gases heated by the stochastic 
thermostat.

\section{The linearized Boltzmann equation}\label{section3}

Let us consider states close to the stationarity, so that we can write
\be\label{deltaf}
\delta f(x_1,t)=f(x_1,t)-f_s(v_1), 
\qquad|\delta f(x_1,t)|\ll f_s(v_1).
\ee
The evolution equation for this function is obtained from the Boltzmann 
equation (\ref{ec.b}) by neglecting the non-linear terms in $\delta f$, 
obtaining
\be
\frac{\partial}{\partial t}\delta f(x_1,t)
+\mathbf{v}_1\cdot\frac{\partial}{\partial\mathbf{r}_1} \delta f(x_1,t)
=K(\mathbf{v}_1)\delta f(x_1,t),
\ee 
where $K(\mathbf{v}_1)$ is a linear operator defined by
\be
K(\mathbf{v}_1)=\sigma^{d-1}\int d\mathbf{v}_2
\bar{T}_0(\mathbf{v}_1,\mathbf{v}_2)(1+P_{12})f_s(v_2)+
\frac{\xi_0^2}{2}\frac{\partial^2}{\partial\mathbf{v}_1^2},
\ee
with $P_{12}$ an operator that interchanges the label $1$ and $2$ in
the function on which it acts. Now, let us introduce a dimensionless space 
variable as 
\be 
\mathbf{l}=\frac{\mathbf{r}}{\ell}, \qquad \ell=(n\sigma^{d-1})^{-1}, 
\ee 
where $\ell$ is proportional to the mean free path, and a dimensionless time
\be
s=\frac{v_s}{\ell}t,
\ee
which is proportional to the number of collisions per particle in the interval
$(0,t)$. It is also convenient to introduce the dimensionless distribution, 
$\delta\chi$ through
\be
\delta f(x,t)=\frac{n}{v_s^d}\delta\chi(\mathbf{l},\mathbf{c},s). 
\ee
The equation for $\delta\chi$ reads
\be\label{ecBL}
\frac{\partial}{\partial s}\delta\chi(\mathbf{l},\mathbf{c}_,s)
=\left[\Lambda(\mathbf{c})
-\mathbf{c}\cdot\frac{\partial}{\partial\mathbf{l}}\right]
\delta\chi(\mathbf{l},\mathbf{c},s). 
\ee
Here we have introduced the homogeneous linearized Boltzmann collision 
operator, $\Lambda$, defined by
\be\label{lBco}
\Lambda(\mathbf{c}_1)=\int d\mathbf{c}_2\bar{T}_0(\mathbf{c}_1,\mathbf{c}_2)
(1+P_{12})\chi_s(c_2)+
\frac{\widetilde{\xi}^2}{2}\frac{\partial^2}{\partial\mathbf{c}_1^2},
\ee
where 
\be\label{barxiSon}
\widetilde{\xi}^2=\frac{\xi_0^2\ell}{v_s^3}
\approx\frac{\pi^{\frac{d-1}{2}}(1-\alpha^2)}{\sqrt{2}d\Gamma(d/2)}
\left(1+\frac{3}{16}a_2^s\right), 
\ee
is the dimensionless amplitude of the noise calculated in the first Sonine 
approximation. 
Since the equation is linear and the linearized Boltzmann collision operator 
does not change the space variable, it is convenient to introduce the Fourier 
component
\be
\delta\chi_{\mathbf{k}}(\mathbf{c},s)=\int d\mathbf{l}
e^{-i\mathbf{k}\cdot\mathbf{l}}\delta\chi(\mathbf{l},\mathbf{c},s). 
\ee
For an infinite system or with periodic boundary conditions, the evolution 
equation for these components is
\be\label{ecBLk}
\frac{\partial}{\partial s}\delta\chi_{\mathbf{k}}(\mathbf{c},s)
=\left[\Lambda(\mathbf{c})
-i\mathbf{k}\cdot\mathbf{c}\right]
\delta\chi_{\mathbf{k}}(\mathbf{c},s). 
\ee
Equation (\ref{ecBL}) or the Fourier counterpart, eq. (\ref{ecBLk}), is the 
so-called linearized Boltzmann equation and it describes the dynamics of any 
small perturbation around the homogeneous stationary state. This provides us with
our starting point for the study of the possibility of a hydrodynamic 
description close to the stationary state. 

The solution of eq. (\ref{ecBLk}) can be written formally as
\be
\delta\chi_{\mathbf{k}}(\mathbf{c},s)=
e^{[\Lambda(\mathbf{c})-i\mathbf{k}\cdot\mathbf{c}]s}
\delta\chi_{\mathbf{k}}(\mathbf{c},0), 
\ee
that shows clearly that the excitations of the gas are determined by the 
spectrum properties of the linear operator 
$\Lambda(\mathbf{c})-i\mathbf{k}\cdot\mathbf{c}$. This suggests the study of 
the eigenvalue problem 
\be\label{eigenP}
[\Lambda(\mathbf{c})-i\mathbf{k}\cdot\mathbf{c}]
\xi_j(\mathbf{k},\mathbf{c})
=\lambda_j(\mathbf{k})\xi_j(\mathbf{k},\mathbf{c}), 
\ee
that is posed in a Hilbert space of functions of $\mathbf{c}$ with scalar 
product
\be\label{scalarHilbert}
\langle g|h\rangle=\int d\mathbf{c}\chi_s^{-1}(c)g^*(\mathbf{c})h(\mathbf{c}),  
\ee
with $g^*$ the complex conjugate of $g$. 
The search for hydrodynamic excitations, which are defined as the ones 
associated to the slowest 
modes, can be carried out by assuming that the modes are analytic in 
$\mathbf{k}$ and looking first for the $\mathbf{k}=\mathbf{0}$ solution of eq. 
(\ref{eigenP}). So, let us consider the homogeneous eigenvalue problem 
\be\label{eigenPH}
\Lambda(\mathbf{c})\xi_j(\mathbf{c})
=\lambda_j\xi_j(\mathbf{c}), 
\ee
where we have introduced the notation 
$\xi_j(\mathbf{k}=\mathbf{0},\mathbf{c})=\xi_j(\mathbf{c})$ and 
$\lambda_j(\mathbf{k}=\mathbf{0})=\lambda_j$. We will now see that the special universal 
solution studied in the previous section allows us to identify a 
family of exact solutions of the linearized Boltzmann equation related to $d+2$ 
modes of the homogeneous linearized collision operator, $\Lambda$. The idea is 
similar in spirit to the one introduced in \cite{bdr03, db03, bd05} to 
calculate the 
hydrodynamic eigenfunctions in the free cooling case. Let us consider the 
family of exact solutions of the homogeneous non-linear Boltzmann equation
\be
f_H(\mathbf{v},t)=\frac{\bar{n}}{\bar{v}_H(t)^d}
\chi\left[\frac{\mathbf{v}-\mathbf{u}}{\bar{v}_H(t)},
\frac{\bar{v}_s}{\bar{v}_H(t)}\right], 
\ee
which are parameterized by the density, $\bar{n}$, the constant velocity flow, 
$\mathbf{u}$, and the thermal velocity, $\bar{v}_H(t)$. The bars on the 
quantities refer to variables that differ from $(n,v_H)$ introduced earlier,  
$\bar{v}_H(t)$ being the thermal 
velocity of the $\beta$-state corresponding to $\bar{n}$. If we consider 
states close to stationarity, the function $\bar{v}_H(t)$ is known and 
given by eq. (\ref{ev.vH}). But then, the family
\be\label{deltafFam}
\delta f(\mathbf{v},t)=\frac{\bar{n}}{\bar{v}_H(t)^d}
\chi\left[\frac{\mathbf{v}-\mathbf{u}}{\bar{v}_H(t)},
\frac{\bar{v}_s}{\bar{v}_H(t)}\right]-f_s(v), 
\ee
to linear order in the fields, has to be a solution of the linearized 
Boltzmann equation. In Appendix \ref{apendiceB} $\delta f$ is calculated to 
linear order, which leads to the corresponding scaled distribution 
\begin{eqnarray}\label{deltaChi}
\delta\chi(\mathbf{c},s)=\frac{\delta n}{n}\left[\chi_s(c)+\frac{1}{3}
\frac{\partial}{\partial\mathbf{c}}\cdot[\mathbf{c}\chi_s(c)]\right]
-\frac{\mathbf{u}}{v_s}\cdot\frac{\partial}{\partial\mathbf{c}}\chi_s(c)
\nonumber\\
-\left[\frac{\delta n}{3n}+\frac{\delta v_H(0)}{v_s}\right]e^{-\gamma s}
\left[\frac{\partial}{\partial\mathbf{c}}\cdot[\mathbf{c}\chi_s(c)]
+\left.\frac{\partial}{\partial\beta}\chi(c,\beta)\right|_{\beta=1}\right], 
\end{eqnarray} 
where $\delta n=\bar{n}-n$ and $\delta v_H(0)=\bar{v}_H(0)-v_s$. The 
functions given by (\ref{deltaChi}) generate a family of solutions of the 
homogeneous linearized Boltzmann equation, that can be seen as 
the superposition of $d+2$ modes. We consequently have identified the following 
eigenfunctions of $\Lambda$ 
\begin{eqnarray}
\xi_1(c)&=&\chi_s(c)+\frac{1}{3}
\frac{\partial}{\partial\mathbf{c}}\cdot[\mathbf{c}\chi_s(c)],
\label{eigen1} \\
\boldsymbol{\xi}_2(\mathbf{c})&=&
-\frac{\partial}{\partial\mathbf{c}}\chi_s(c), \label{eigen2} \\
\xi_3(c)&=&-\frac{\partial}{\partial\mathbf{c}}\cdot[\mathbf{c}\chi_s(c)]
-\left.\frac{\partial}{\partial\beta}\chi(c,\beta)\right|_{\beta=1}, 
\label{eigen3} 
\end{eqnarray}
with the corresponding eigenvalues
\be
\lambda_1=\lambda_2=0,  
\qquad \lambda_3=-\gamma, 
\ee
where the null eigenvalue is $(d+1)$-fold degenerate.  Moreover, 
eq. (\ref{deltaChi}) can be rewritten as
\be
\delta\chi(\mathbf{c},s)=\sum_{j=1}^{3}e^{\lambda_js}
\langle\bar{\xi}_j(\mathbf{c})|\delta\chi(\mathbf{c},0)\rangle
\xi_j(\mathbf{c}),  
\ee
with the following definitions of the functions $\bar{\xi}_i$
\be\label{barxidef}
\bar{\xi}_1(\mathbf{c})=\chi_s(c),
\qquad\bar{\boldsymbol{\xi}}_2(\mathbf{c})=\chi_s(c)\mathbf{c}
\qquad\bar{\xi}_3=\chi_s(c)\left(\frac{c^2}{d}-\frac{1}{6}\right). 
\ee
These functions are just the linear combinations of the the first two velocity 
moments, $\{1,\mathbf{c},c^2\}$, normalized to enforce
\be\label{biorth.cond}
\langle\bar{\xi}_i|\xi_j\rangle=\delta_{ij}, \qquad i,j=1,2,3. 
\ee
In fact, the functions $\xi_1$ and $\boldsymbol{\xi}_2$ have been previously 
identified in \cite{gmt09, mgt09, mg11}, where they were used to study the 
fluctuations of quantities like the total energy in the stationary state. There, it was also 
proven that $\bar{\xi}_1$ and $\bar{\boldsymbol{\xi}}_2$ are eigenfunctions of 
the adjoint operator of the linearized collision operator, $\Lambda^+$, 
associated to the null eigenvalue. This follows from the conservation of the number 
of particles and momentum:
\be
\int d\mathbf{c}\Lambda(\mathbf{c})h(\mathbf{c})
=\int d\mathbf{c} c_i\Lambda(\mathbf{c})h(\mathbf{c})=0.
\ee
Then, if we assume that eigenfunctions of $\Lambda$, 
$\{\xi_i\}_{i=1}^{\infty}$, form a complete set, in the expansion 
\be
f(\mathbf{c})=\sum_{j=1}^{\infty}C_j\xi_j(\mathbf{c}), 
\ee
we have $C_1=\langle\bar{\xi}_1|f\rangle$ and 
$\mathbf{C}_2=\langle\bar{\boldsymbol{\xi}}_2|f\rangle$. With the aid of eq. 
(\ref{deltaChi}), we are able to identify a new eigenfunction, $\xi_3$, that, 
as seen in (\ref{eigen3}), depends on the derivative with respect to $\beta$ of 
the universal state $\chi$ in the stationary regime. Let us remark that 
$\bar{\xi}_3$ is not an eigenfunction of $\Lambda^+$. We only have 
$C_3=\langle\bar{\xi}_3|f\rangle$ if the function $f$ belongs to the space 
generated by the set $\{\xi_i\}_{i=1}^3$. 

In the elastic case, i.e. $\alpha=1$ and $\xi_0^2=0$, the spectrum of 
$\Lambda(\mathbf{k},\mathbf{c})$ has been analyzed in detail and it is known that 
the modes associated to the locally conserved quantities, the hydrodynamic 
modes, are the slowest ones in 
the $k\to 0$ limit, where the respective eigenvalues vanish \cite{mcLennan}. 
Furthermore, it is known that the 
spectrum is analytic in $k$  around $k=0$, and that there is scale separation, 
i.e. the hydrodynamic modes are isolated from the rest of modes. 
For the inelastic linearized collision 
operator,  no such result is available. We have nevertheless shown 
that $d+1$ of the $d+2$ identified modes are associated to the vanishing 
eigenvalue, and the remaining mode is associated to $\gamma$, which itself vanishes in 
the elastic limit. We therefore expect that a similar property will hold with 
the identified modes --at least close to the elastic limit-- and these modes will
henceforth be coined `hydrodynamic'. In the following, we will \emph{assume} 
that they are the slowest ones and that they are analytic. Under this proviso, 
the asymptotic behavior of the one particle distribution function 
for $k \ll 1$, $s \gg1$ is
\be\label{chiHydro}
\delta\chi_{\mathbf{k}}(\mathbf{c},s)\approx\sum_{j=1}^{d+2}K_j
e^{\lambda_j(k)s}\xi_j(\mathbf{c}), 
\ee
where the $\{K_j\}_{j=1}^{d+2}$ depend on the initial condition. Note that, as 
seen in eqs. (\ref{deltafFam}) and (\ref{deltaChi}), the hydrodynamic 
eigenfunctions 
given by eqs. (\ref{eigen1})-(\ref{eigen3}), expand the subspace of functions 
generated by the difference of a local $\beta$-state with the stationary state. 
Then, eq. (\ref{chiHydro}) can be rewritten in the original variables as
\be\label{localBetaState}
\delta f(\mathbf{r},\mathbf{v},t)
\approx\frac{n(\mathbf{r},t)}{v(\mathbf{r},t)^d}
\chi\left[\frac{\mathbf{v}-\mathbf{u}(\mathbf{r},t)}{v(\mathbf{r},t)},
\frac{\bar{v}_s[n(\mathbf{r},t)]}{v(\mathbf{r},t)} \right]-f_s(v). 
\ee
This shows that for small gradients (or equivalently for $k \ll 1$)
and in the long time limit, i.e. in the time in which the 
non-hydrodynamic modes have decayed, all the time dependence in the 
distribution function is through the hydrodynamic fields. 
Moreover, the 
distribution function takes the form of a local $\beta$-state distribution, 
which plays, in this context, the role of a reference state. To 
evaluate explicitly the time evolution of the fields, it is necessary to 
calculate $\{\lambda_j(k)\}_{j=1}^{d+2}$. If we assume that they are analytic in 
$k$, this can be carry out studying the eigenvalue problem, eq. (\ref{eigenP}),  
by standard perturbation theory, as was performed in 
\cite{db03, bd05} for the free cooling case. In the next section, we will 
evaluate the eigenvalues 
to Navier-Stokes order, i.e. $k^2$ order, but by a different method, deriving 
the evolution equations for the linear deviations of the hydrodynamic fields.

\section{Linear hydrodynamics around the stationary state}
\label{section4}

The objective of this section is to derive evolution equations for the 
deviations of the hydrodynamic fields around its stationary values at 
Navier-Stokes order. The analysis of these equations will clarify 
the differences and analogies with respect to the elastic case. 
The hydrodynamic eigenvalues, $\{\lambda_j(k)\}_{j=1}^{d+2}$, will be obtained by 
identifying the asymptotic behavior of the solutions in the proper limit. Let 
us start writing the deviations of the hydrodynamic fields as 
\begin{eqnarray}
\rho(\mathbf{l},s)&\equiv&\frac{n(\mathbf{r},t)-n}{n}
=\int d\mathbf{c}\delta\chi(\mathbf{l},\mathbf{c},s),  \\
\mathbf{w}(\mathbf{l},s)&\equiv&\frac{\mathbf{u}(\mathbf{r},t)}{v_s} 
=\int d\mathbf{c}\mathbf{c}\delta\chi(\mathbf{l},\mathbf{c},s), \\
\theta(\mathbf{l},s)&\equiv&\frac{T(\mathbf{r},t)-T_s}{T_s}
=\int d\mathbf{c}\left(\frac{2c^2}{d}-1\right)
\delta\chi(\mathbf{l},\mathbf{c},s), 
\end{eqnarray}
that can be expressed in terms of the scalar products of the distribution 
function with the functions $\{\bar{\xi}_j\}_{j=1}^{d+2}$. In Fourier space, 
they read
\begin{eqnarray}\label{comp1}
\langle\bar{\xi}_1(c)|\delta\chi_{\mathbf{k}}(\mathbf{c},s)\rangle
&=&\rho_{\mathbf{k}}(s), \\
\label{comp2}
\langle\bar{\boldsymbol{\xi}}_2(\mathbf{c})|
\delta\chi_{\mathbf{k}}(\mathbf{c},s)\rangle
&=&\mathbf{w}_{\mathbf{k}}(s), \\
\label{comp3}
\langle\bar{\xi}_3(c)|\delta\chi_{\mathbf{k}}(\mathbf{c},s)\rangle
&=&\frac{1}{2}\theta_{\mathbf{k}}(s)+\frac{1}{3}\rho_{\mathbf{k}}(s). 
\end{eqnarray}
Now, let us define the relevant projector in the hydrodynamic subspace
\be\label{projP}
\mathcal{P}h(\mathbf{c})=\sum_{j=1}^{d+2}
\langle\bar{\xi}_j(\mathbf{c})|h(\mathbf{c})\rangle\xi_j(\mathbf{c}), 
\ee
and also the orthogonal one
\be
\mathcal{Q}=\mathcal{I}-\mathcal{P}, 
\ee
where $\mathcal{I}$ is the identity operator. As alluded to above, 
$\langle\bar{\xi}_3|h\rangle$ is not the actual component of $h$ into $\xi_3$ 
but, still, eq. (\ref{projP}) defines a projector as 
$\mathcal{P}^2=\mathcal{I}$ (we also have $\mathcal{Q}^2=\mathcal{I}$). 

If we apply the projectors $\mathcal{P}$ and $\mathcal{Q}$ to the linearized 
Boltzmann equation, eq. (\ref{ecBLk}), we obtain the following set of coupled 
equations
\begin{eqnarray}
\left[\frac{\partial}{\partial s}-
\mathcal{P}[\Lambda(\mathbf{c})-i\mathbf{k}\cdot\mathbf{c}]
\mathcal{P}\right]\mathcal{P}\delta\chi_{\mathbf{k}}(\mathbf{c},s)
&=&\left[\mathcal{P}\Lambda(\mathbf{c})
-\mathcal{P}i\mathbf{k}\cdot\mathbf{c}\right]\mathcal{Q}
\delta\chi_{\mathbf{k}}(\mathbf{c},s), \label{ec.Pdchi}\\
\left[\frac{\partial}{\partial s}
-\mathcal{Q}[\Lambda(\mathbf{c})-i\mathbf{k}\cdot\mathbf{c}]
\mathcal{Q}\right]\mathcal{Q}\delta\chi_{\mathbf{k}}(\mathbf{c},s)
&=&-\mathcal{Q}i\mathbf{k}\cdot\mathbf{c}\mathcal{P}
\delta\chi_{\mathbf{k}}(\mathbf{c},s), \label{ec.Qdchi}
\end{eqnarray}
where we have used that $\mathcal{Q}\Lambda\mathcal{P}=0$. However, 
$\mathcal{P}\Lambda\mathcal{Q}\ne 0$ because $\bar{\xi}_3$ is not a left 
eigenfunction of $\Lambda$. In fact, we have 
\be
\mathcal{P}\Lambda(\mathbf{c})\mathcal{Q}h(\mathbf{c})=\xi_3(c)
\langle\bar{\xi}_3(c)|\Lambda(\mathbf{c})\mathcal{Q}h(\mathbf{c})\rangle. 
\ee
The $d+2$ components of eq. (\ref{ec.Pdchi}) are the evolution equations for 
the hydrodynamic fields 
\begin{eqnarray}
&&\frac{\partial}{\partial s}\rho_{\mathbf{k}}(s)
+i\mathbf{k}\cdot\mathbf{w}_{\mathbf{k}}(s)=0, \\
&&\frac{\partial}{\partial s}\mathbf{w}_{\mathbf{k}}(s)
+\frac{i}{2}\mathbf{k}\left[\rho_{\mathbf{k}}(s)+\theta_{\mathbf{k}}(s)\right]
+i\mathbf{k}\cdot\boldsymbol{\overleftrightarrow\Pi}_{\mathbf{k}}(s)=\mathbf{0}, 
\\
&&\frac{\partial}{\partial s}\theta_{\mathbf{k}}(s)
+\gamma\left[\frac{2}{3}\rho_{\mathbf{k}}(s)+\theta_{\mathbf{k}}(s)\right]
+\frac{2i}{d}\mathbf{k}\cdot\mathbf{w}_{\mathbf{k}}(s)
+\frac{2i}{d}\mathbf{k}\cdot\boldsymbol{\phi}_{\mathbf{k}}(s)
=\delta\zeta_{\mathbf{k}}(s), 
\end{eqnarray}
where we have introduced the pressure tensor and heat flux
\begin{equation}\label{fluxes}
\boldsymbol{\overleftrightarrow\Pi}_{\mathbf{k}}(s)=\int d\mathbf{c}
\boldsymbol{\overleftrightarrow\Delta}(\mathbf{c})
\mathcal{Q}\delta\chi_{\mathbf{k}}(\mathbf{c},s), \qquad 
\boldsymbol{\phi}_{\mathbf{k}}(s)=\int d\mathbf{c}\boldsymbol{\Sigma}(\mathbf{c})
\mathcal{Q}\delta\chi_{\mathbf{k}}(\mathbf{c},s), 
\end{equation}
with
\begin{equation}
\Delta_{jp}(\mathbf{c})=c_jc_p-\frac{c^2}{d}\delta_{jp}, \qquad 
\Sigma_j(\mathbf{c})=\left(c^2-\frac{d+2}{2}\right)c_j, 
\end{equation}
and the deviation of the cooling rate
\begin{equation}\label{coolingRate.def}
\delta\zeta_{\mathbf{k}}(s)=\int d\mathbf{c}\frac{2c^2}{d}\Lambda(\mathbf{c})
\mathcal{Q}\delta\chi_{\mathbf{k}}(\mathbf{c},s). 
\end{equation}
Let us note that the operator $\mathcal{Q}$ can be skipped in eq. 
(\ref{fluxes}) because, due to symmetry properties,  we have
\be
\langle\chi(c)\Delta_{jp}(\mathbf{c})|\xi_{\beta}(\mathbf{c})\rangle=
\langle\chi(c)\Sigma_{j}(\mathbf{c})|\xi_{\beta}(\mathbf{c})\rangle=0 \quad
j,p=1,\dots,d, \quad \beta=1, \dots d+2. 
\ee

To evaluate the hydrodynamic equations to $k^2$ order, we need the fluxes 
given by eq. (\ref{fluxes}), and consequently $\delta\chi_{\mathbf{k}}$ to first 
order in $k$. This is evaluated in Appendix \ref{apendiceC} for an initial 
condition of the form $\mathcal{Q}\delta\chi_{\mathbf{k}}(\mathbf{c},0)=0$ 
and neglecting all the $k$ contributions in the kinetic modes, yielding
\be\label{Qdeltachi1}
\mathcal{Q}\delta\chi_{\mathbf{k}}(\mathbf{c},s)\approx -\int_0^sds'
e^{\mathcal{Q}\Lambda(\mathbf{c})\mathcal{Q}(s-s')}\mathcal{Q}
i\mathbf{k}\cdot\mathbf{c}\mathcal{P}\delta\chi_{\mathbf{k}}(\mathbf{c},s'). 
\ee
Note that, as the hydrodynamic eigenfunctions expand the subspace of functions 
generated by the difference of a local $\beta$-state with the stationary state, 
the condition $\mathcal{Q}\delta\chi_{\mathbf{k}}(\mathbf{c},0)=0$ represents an 
initial condition of the local $\beta$-state form. 
When (\ref{Qdeltachi1}) is inserted in (\ref{fluxes}), taking into account the 
symmetries of the system as discussed in Appendix \ref{apendiceC}, the following 
expressions for the fluxes are obtained 
\be
\Pi_{\mathbf{k}, jp}(s)=-i\int_0^sds'G_{xy}(s-s')
\left[k_jw_{\mathbf{k},p}(s')+k_pw_{\mathbf{k},j}(s')
-\frac{2}{d}\delta_{jp}\mathbf{k}\cdot\mathbf{w}_{\mathbf{k}}(s')\right], 
\ee
and
\be\label{phiExp}
\phi_{\mathbf{k},j}(s)=-ik_j\int_0^sds'
\left\{\rho_{\mathbf{k}}(s')\left[H_1(s-s')+\frac{1}{3}H_3(s-s')\right]
+\frac{1}{2}\theta_{\mathbf{k}}(s')H_3(s-s')\right\}, 
\ee
where we have introduced the ``correlation'' functions
\be
G_{xy}(s)=\int d\mathbf{c}\Delta_{xy}(\mathbf{c})e^{\Lambda(\mathbf{c})s}
c_x\xi_{2,y}(\mathbf{c}), 
\ee
and 
\be
H_j(s)=\int d\mathbf{c}\Sigma_{x}(\mathbf{c})e^{\Lambda(\mathbf{c})s}
c_x\xi_j(\mathbf{c}), \qquad j=1,3. 
\ee
It is important to remark that, as $\Delta_{xy}$ and $\Sigma_x$ are orthogonal 
to the hydrodynamic modes, the functions $G_{xy}(s)$ and $H_j(s)$ decay with the 
kinetic modes. 
Under the same hypothesis as for the fluxes, the cooling rate, 
$\delta\zeta_{\mathbf{k}}$, is evaluated in Appendix \ref{apendiceD} to $k^2$ 
order, with the result 
\begin{eqnarray}
&&\delta\zeta_{\mathbf{k}}(s)=-2i\int_0^sds'
\mathbf{k}\cdot\mathbf{w}_{\mathbf{k}}(s')Z(s-s')\nonumber\\
&&-2k^2\int_0^sds'
\left\{\rho_{\mathbf{k}}(s')\left[Z_1(s-s')+\frac{1}{3}Z_3(s-s')\right]
+\frac{1}{2}\theta_{\mathbf{k}}(s')Z_3(s-s')\right\}. \nonumber\\
\end{eqnarray}
We have introduced here the functions
\be
Z(s)=\langle\bar{\xi}_3(\mathbf{c})|[\Lambda(\mathbf{c})-\lambda_3]
e^{\Lambda(\mathbf{c})s}c_x\xi_{2,x}(\mathbf{c})\rangle, 
\ee
and
\be
Z_j(s)=\langle\bar{\xi}_3(\mathbf{c})|[\Lambda(\mathbf{c})-\lambda_3]\int_0^s
ds'e^{\Lambda(\mathbf{c})(s-s')}c_x\mathcal{Q}e^{\Lambda(\mathbf{c})s'}
c_x\xi_{j}(\mathbf{c})\rangle, \quad j=1,3, 
\ee
that, due to the bi-orthogonality condition, eq. (\ref{biorth.cond}), also decay 
with the kinetic modes. Note that, at variance with the free cooling 
case \cite{bdks98}, there is here a first order in $k$ contribution to the 
cooling rate. 

It also proves convenient to introduce the parallel and transversal components of 
the velocity 
\be
w_{\mathbf{k},||}(s)=\hat{\mathbf{k}}\cdot\mathbf{w}_{\mathbf{k}}(s), \qquad 
w_{\mathbf{k}, \perp}^{(j)}=\hat{\mathbf{k}}_{\perp}^{(j)}\cdot
\mathbf{w}_{\mathbf{k}}(s), \quad j=1,\dots,d-1, 
\ee
where $\hat{\mathbf{k}}=\mathbf{k}/k$ is a unit vector in the direction of 
$\mathbf{k}$ and $\{\hat{\mathbf{k}}_{\perp}^{(j)}\}_{j=1}^{d-1}$ is an orthogonal 
basis of the subspace orthogonal to $\mathbf{k}$. In terms of these components, 
the hydrodynamic equations are 
\begin{eqnarray}\label{eH1}
&&\frac{\partial}{\partial s}\rho_{\mathbf{k}}(s)
+ikw_{\mathbf{k}, ||}(s)=0, \\ 
\label{eH2}
&&\frac{\partial}{\partial s}w_{\mathbf{k},\perp}^{(j)}(s)
+k^2\int_0^sds'G_{xy}(s-s')w_{\mathbf{k}, \perp}^{(j)}(s')=0, \quad j=1,\dots,d-1\\
\label{eH3}
&&\frac{\partial}{\partial s}w_{\mathbf{k}, ||}(s)
+\frac{i}{2}k\left[\rho_{\mathbf{k}}(s)+\theta_{\mathbf{k}}(s)\right]
+2\frac{d-1}{d}k^2\int_0^sds'G_{xy}(s-s') w_{\mathbf{k}, ||}(s')=0, \\
\label{eH4}
&&\frac{\partial}{\partial s}\theta_{\mathbf{k}}(s)
+\gamma\left[\frac{2}{3}\rho_{\mathbf{k}}(s)+\theta_{\mathbf{k}}(s)\right]
+\frac{2i}{d}kw_{\mathbf{k},||}(s)+2ik\int_0^sds'Z(s-s') w_{\mathbf{k}, ||}(s') 
\nonumber \\
&&+\frac{2}{d}k^2\int_0^sds'[G_1(s-s')\rho_{\mathbf{k}}(s')
+G_3(s-s')\theta_{\mathbf{k}}(s')]
=0, 
\end{eqnarray}
where we have introduced the new correlation functions
\be
G_1(s)=H_1(s)+\frac{1}{3}H_3(s)+dZ_1(s)+\frac{d}{3}Z_3(s), \qquad
G_3(s)=\frac{1}{2}H_3(s)+\frac{d}{2}Z_3(s). 
\ee 
It is noteworthy that, apart from the assumption of an initial 
condition in the hydrodynamic subspace, the only approximations made in the 
derivation of eqs. (\ref{eH1})-(\ref{eH4}) are the expansion to second 
order in the gradients of the hydrodynamic fields and the neglect of the $k$ 
contribution in the memory kernels that all decay with the kinetic modes. Of 
course, the kernels are for the moment unknown, but we will see later that 
they can be calculated approximately. Before doing so, we 
evaluate the asymptotic behavior of eqs. (\ref{eH1})-(\ref{eH4}) in the 
hydrodynamic limit (under which eq. (\ref{chiHydro}) was derived). 

Due to the 
structure of the equations, it is convenient to introduce the Laplace 
transforms of the fields
\be
\bar{f}(z)=\int_0^\infty dse^{-zs}f(s), 
\qquad f(s)=\frac{1}{2\pi i}\int_{c-i\infty}^{c+i\infty} dze^{zs}\bar{f}(z),  
\ee
where $c$ is bigger than the real part of all the poles of $\bar{f}$. In the 
Laplace space the convolutions transform into products  
and Eqs. (\ref{eH1})-(\ref{eH4}) become

\be\label{eHLw}
z\bar{w}_{\mathbf{k},\perp}^{(j)}(z)
+k^2\bar{G}_{xy}(z)\bar{w}_{\mathbf{k},\perp}^{(j)}(z)
=w_{\mathbf{k},\perp}^{(j)}(0), \qquad j=1,\dots,d-1. 
\ee
\be\label{eHLcouple}
[zI+A(k,z)]\left(\begin{array}{c}
\bar{\rho}_{\mathbf{k}}(z)\\
\bar{w}_{\mathbf{k}, ||}(z)\\
\bar{\theta}_{\mathbf{k}}(z)\end{array}\right)=
\left(\begin{array}{c}\rho_{\mathbf{k}}(0)\\w_{\mathbf{k}, ||}(0)\\
\theta_{\mathbf{k}}(0)\end{array}\right), 
\ee
where we have introduced the matrix
\be\label{matrixA}
A(k,z)=\left(\begin{array}{ccc} 0 & ik & 0 \\
\frac{i}{2}k & 2\frac{d-1}{d}\bar{G}_{xy}(z)k^2 & \frac{i}{2}k \\
\frac{2}{3}\gamma+\frac{2}{d}\bar{G}_1(z)k^2 & i\bar{q}(z)k & 
\gamma+\frac{2}{d}\bar{G}_3(z)k^2\end{array}\right), 
\ee
with
\be
\bar{q}(z)=\frac{2}{d}[1+d\bar{Z}(z)]. 
\ee
The equation for the transverse velocity, eq. (\ref{eHLw}), is analyzed in 
detail in Appendix \ref{apendiceE}. Assuming that $G_{xy}(s)$ is a linear 
combination of kinetic modes (as is expected to be), we obtain in the 
hydrodynamic limit
\be\label{vtransInt}
w_{\mathbf{k},\perp}^{(j)}(s)\approx 
w_{\mathbf{k},\perp}^{(j)}(0)e^{-\eta k^2s}, 
\ee
where we have introduced the shear viscosity
\be\label{viscosi}
\eta=\int_0^\infty dsG_{xy}(s). 
\ee
In the same limit, the solution of the coupled hydrodynamic equations, eq. 
(\ref{eHLcouple}), is analyzed in Appendix \ref{apendiceF}, obtaining 
\be\label{ceSol}
|y(\mathbf{k},s)\rangle\approx\sum_{\beta=1}^{3}\langle\phi_{\beta}
|y(\mathbf{k},0)\rangle e^{\lambda_{\beta}(k)s} |\psi_{\beta}^{(0)}\rangle, 
\ee
where we have introduced the notation
\be
|y(\mathbf{k},s)\rangle\equiv\left(\begin{array}{c}
\rho_{\mathbf{k}}(s)\\
w_{\mathbf{k}, ||}(s)\\
\theta_{\mathbf{k}}(s)\end{array}\right). 
\ee
The functions $\{|\psi_{\beta}^{(0)}\rangle\}_{\beta=1}^3$ are the zeroth order in 
$k$ contribution to the expansion of the eigenfunctions, 
$|\psi_{\beta}(k,z)\rangle$, of the matrix $A(k,z)$
\be
|\psi_{\beta}(k,z)\rangle=|\psi_{\beta}^{(0)}(z)\rangle
+k|\psi_{\beta}^{(1)}(z)\rangle+k^2|\psi_{\beta}^{(2)}(z)\rangle+\dots,
\ee
with $A(k,z)|\psi_{\beta}(k,z)\rangle=a_{\beta}(k,z)|\psi_{\beta}(k,z)\rangle$. 
They are calculated in Appendix \ref{apendiceF} obtaining
\be
|\psi_1^{(0)}\rangle=\left(\begin{array}{c}
\ -6 \\ \sqrt{6} \\ 4 \end{array}\right), \qquad
|\psi_2^{(0)}\rangle=\left(\begin{array}{c}
\ 6 \\ \sqrt{6} \\ -4 \end{array}\right), \qquad
|\psi_3^{(0)}\rangle=\left(\begin{array}{c}
\ 0 \\ 0 \\ 1 \end{array}\right), \qquad
\ee
where we see that they do not depend on $z$. The other set of functions, 
$\{\langle\phi_{\beta}|\}_{\beta=1}^3$, is the bi-orthogonal set
\be
\langle\phi_1|=\left(-\frac{1}{12}, \frac{1}{2\sqrt{6}}, 0 \right), \qquad
\langle\phi_2|=\left(\frac{1}{12}, \frac{1}{2\sqrt{6}}, 0 \right), \qquad
\langle\phi_3|=\left(\frac{2}{3}, 0, 1 \right), 
\ee
that is constructed to have 
$\langle\phi_{\beta}|\psi_{\beta'}^{(0)}\rangle=\delta_{\beta,\beta'}$. The scalar 
product $\langle u|v\rangle$ between two vectors $|u\rangle$ and $|v\rangle$ is 
the usual Euclidean scalar product. There is no confusion with the one 
introduced in eq. (\ref{scalarHilbert}) because they appear in different 
contexts. The explicit expressions of the hydrodynamic eigenvalues, 
$\{\lambda_\beta(k)\}_{\beta=1}^3$, to $k^2$ order are
\begin{eqnarray}\label{lambda1approx}
\lambda_1(k)&=&+\frac{i}{\sqrt{6}}k-\left[\frac{d-1}{d}\eta
+\frac{1}{4\gamma}\left(\frac{2}{3}+q_0\right)\right]k^2, \\
\label{lambda2approx}
\lambda_2(k)&=&-\frac{i}{\sqrt{6}}k-\left[\frac{d-1}{d}\eta
+\frac{1}{4\gamma}\left(\frac{2}{3}+q_0\right)\right]k^2, \\
\label{lambda3approx}
\lambda_3(k)&=&-\gamma+\left[\frac{1}{3\gamma}+\frac{1}{2\gamma}q_w
-\frac{2}{d}\left(\widetilde{\kappa}+\zeta_{\theta}\right)\right]k^2, 
\end{eqnarray}
where the shear viscosity, $\eta$, is given in (\ref{viscosi}), and we have 
introduced the heat conductivity, $\widetilde{\kappa}$, as
\be\label{kappaExp}
\widetilde{\kappa}=\frac{1}{2}\int_0^{\infty}dse^{\gamma s}H_3(s). 
\ee
The expressions for the other transport coefficients are
\begin{eqnarray}\label{q0Exp}
q_0&=&\frac{2}{d}+2\int_0^{\infty}dsZ(s), \\
\label{qwExp}
q_w&=&\frac{2}{d}+2\int_0^{\infty}dse^{\gamma s}Z(s), \\
\label{ztExp}
\zeta_{\theta}&=&\frac{d}{2}\int_0^{\infty}dse^{\gamma s}Z_3(s). 
\end{eqnarray}

Eqs. (\ref{vtransInt}) and (\ref{ceSol}) with the expressions for the transport 
coefficients, eqs. (\ref{viscosi}), (\ref{kappaExp})-(\ref{ztExp}) are the main 
results of the paper. It is important to remark that the equations are valid in 
the linear regime close to the stationary state, but the transport 
coefficients depend on the structure of the non-stationary $\beta$-state 
through the eigenvalue $\gamma$ and the third eigenfunction. Actually, similar 
effects are shown to be essential to understand the dynamics of a homogeneous 
perturbation of the temperature close to the steady uniform shear flow for 
granular gases \cite{bmg12}. Let us also note that, although there is a 
coupling between the heat flux and the density as it is seen in eq. 
(\ref{phiExp}), this 
coupling is not reflected at the level of Navier-Stokes linear hydrodynamic 
(there is no contribution from $H_1(s)$ in the transport coefficients). This is 
also the case in the free cooling case where the hydrodynamic eigenvalues to 
$k^2$ order do not depend on the diffusive conductivity \cite{db03, bd05}. 
Moreover, 
the memory kernel $Z(s)$ appears in $q_0$ and $q_w$ weighted in different ways, 
reflecting the non-Markovian character of eqs. (\ref{eH1})-(\ref{eH4}). 
Nevertheless, as we shall see in the remainder, these effects are expected to 
be small. We also emphasize that the viscosity and heat 
conductivity have been calculated in \cite{Bada2012} applying the 
Chapmann-Enskog expansion to the inelastic Enskog equation, obtaining 
equivalent expressions for both transport coefficients in the low-density 
limit. 

Our goal is now to calculate all the correlations functions, 
$G_{xy}(s)$, $G_{1}(s)$, $G_{3}(s)$ and $Z(s)$, that appear in the hydrodynamic 
equations, eqs. (\ref{eH1})-(\ref{eH4}), in an approximate way. With this, we 
will be able to obtain explicit formulas for all the transport coefficients. 
The idea is reminiscent of that used for free-cooling systems 
\cite{bmg11} and consists in treating the functions $\bar{\xi}_3(c)$, 
$\chi_s(c)\Delta_{jp}(\mathbf{c})$, and $\chi_s(c)\Sigma_{j}(\mathbf{c})$ as if 
they 
were eigenfunctions of the adjoint linearized Boltzmann operator, $\Lambda^{+}$ 
(the adjoint is taken with the scalar product of eq. (\ref{scalarHilbert})). 
That is, we assume
\be
\Lambda^{+}(\mathbf{c})\bar{\xi}_3(c)\approx\lambda_3\bar{\xi}_3(c), 
\ee
and 
\be\label{aprox2}
\Lambda^{+}(\mathbf{c})\chi_s(c)\Delta_{jp}(\mathbf{c})\approx
\lambda_{NH}^{(1)}\chi_s(c)\Delta_{jp}(\mathbf{c}), \quad
\Lambda^{+}(\mathbf{c})\chi_s(c)\Sigma_{j}(\mathbf{c})\approx
\lambda_{NH}^{(2)}\chi_s(c)\Sigma_{j}(\mathbf{c}),
\ee
where $\lambda_{NH}^{(1)}$ and $\lambda_{NH}^{(2)}$ are two non-hydrodynamic 
(kinetic) eigenvalues that have to be calculated consistently. Within this 
approximation we trivially have
\be
Z(s)\approx 0, \quad Z_j(s)\approx 0, \quad j=1,3, 
\ee
so that
\be
q_0\approx \frac{2}{d}, \qquad q_w\approx \frac{2}{d}, 
\qquad \zeta_{\theta}\approx 0, 
\ee
and 
\be
G_1(s)\approx H_1(s)+\frac{1}{3}H_3(s), \qquad G_3(s)\approx \frac{1}{2}H_3(s). 
\ee
Assuming the property (\ref{aprox2}) holds, the functions $G_{xy}(s)$, $H_1(s)$, and 
$H_3(s)$ follow by straightforward calculations:
\be
G_{xy}(s)=\frac{1}{2}e^{\lambda_{NH}^{(1)}s}, \qquad
H_1(s)=-\frac{(d+2)(2+a_2^s)}{12}e^{\lambda_{NH}^{(2)}s}, 
\ee
\be
H_3(s)=\left[(d+2)\left(\frac{1}{2}+a_2^s\right)
-\frac{d+2}{4}\left.\frac{da_2(\beta)}{d\beta}\right|_{\beta=1}\right]
e^{\lambda_{NH}^{(2)}s}, 
\ee
where the eigenvalues $\lambda_{NH}^{(1)}$ and $\lambda_{NH}^{(2)}$ are calculated 
in Appendix \ref{apendiceG} in the approximation of eq. (\ref{aprox2}). With 
these functions, the transport coefficient are easily calculated 
\be\label{tcPrediction}
\eta=\frac{1}{2|\lambda_{NH}^{(1)}|}, 
\qquad \widetilde{\kappa}=\frac{1}{2\left(|\lambda_{NH}^{(2)}|-\gamma\right)}
\left[(d+2)\left(\frac{1}{2}+a_2^s\right)
-\frac{d+2}{4}\left.\frac{da_2(\beta)}{d\beta}\right|_{\beta=1}\right]. 
\ee

\begin{figure}
\begin{minipage}{0.48\linewidth}
\begin{center}
\includegraphics[angle=0,width=0.9\linewidth]{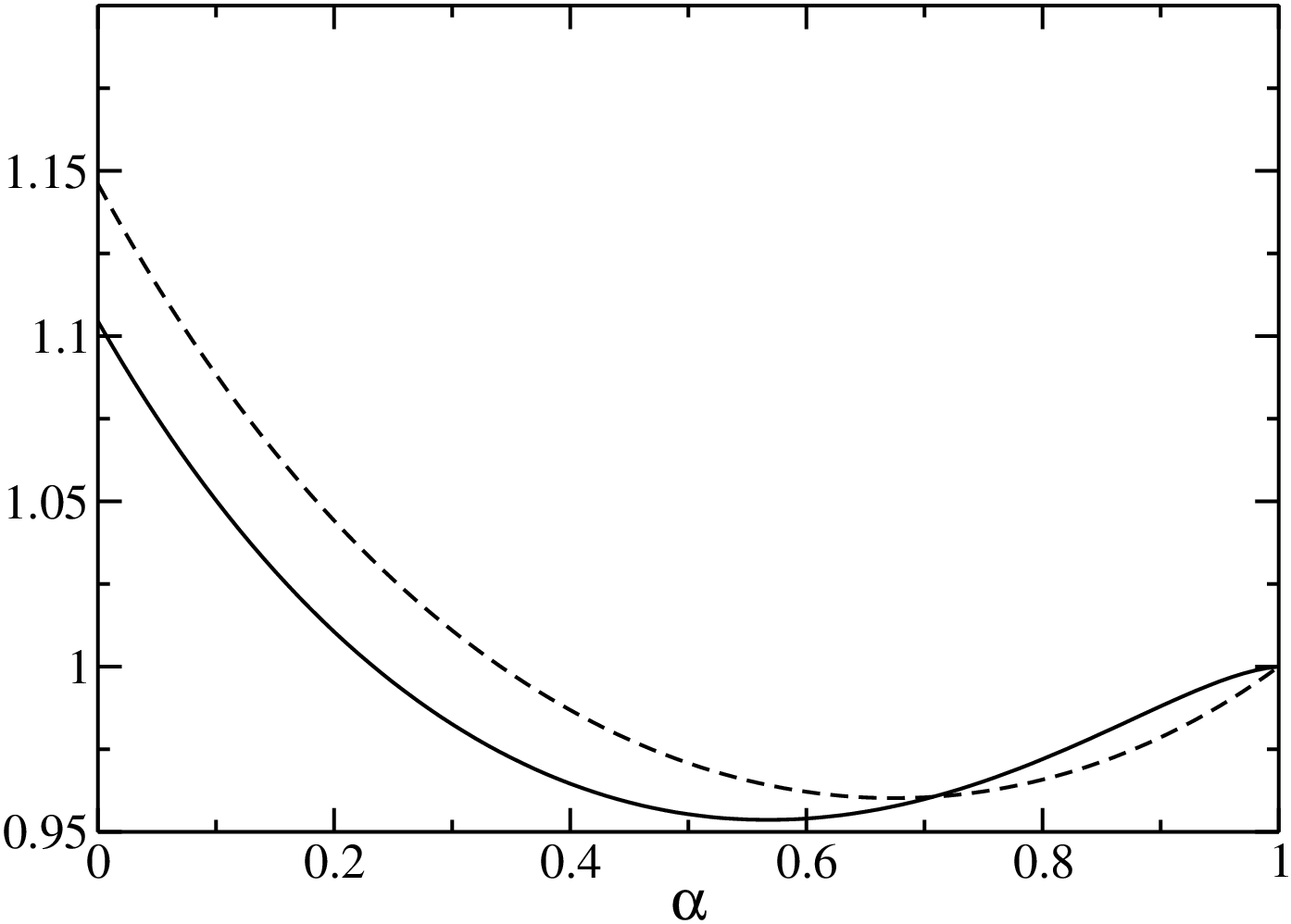}
\end{center}
\end{minipage}
\begin{minipage}{0.48\linewidth}
\begin{center}
\includegraphics[angle=0,width=0.9\linewidth]{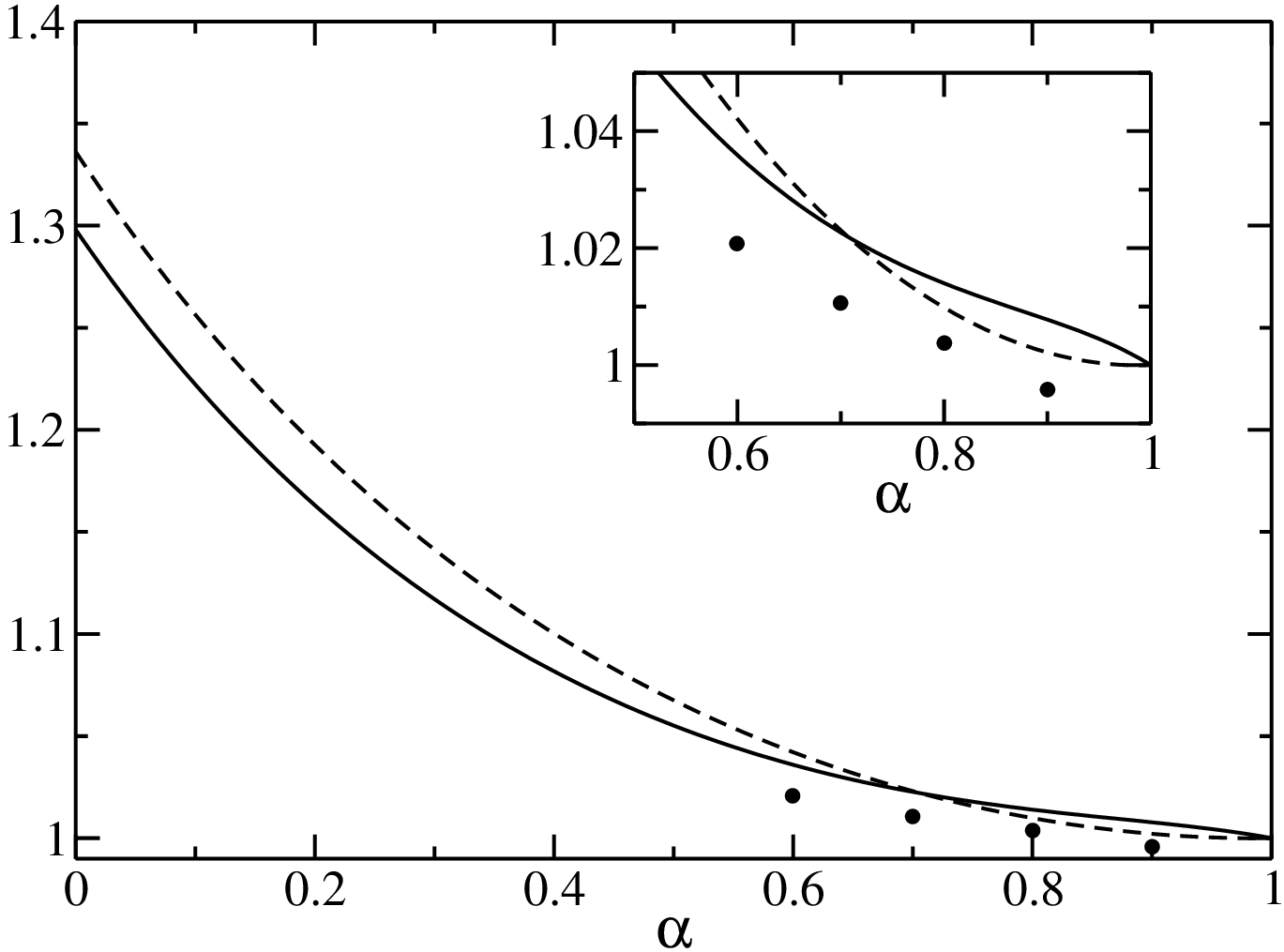}
\end{center}
\end{minipage}
\caption{Reduced viscosity of a two-dimensional system (left) and 
three-dimensional system (right) as a function of the inelasticity. The 
solid line is the theoretical prediction given by eq. (\ref{tcPrediction}), 
the dashed line is the theoretical prediction of \cite{gm02}, and the dots are 
the simulation results of \cite{gm02}. }\label{fig1}
\end{figure}

\begin{figure}[h]
\begin{minipage}{0.48\linewidth}
\begin{center}
\includegraphics[angle=0,width=0.9\linewidth]{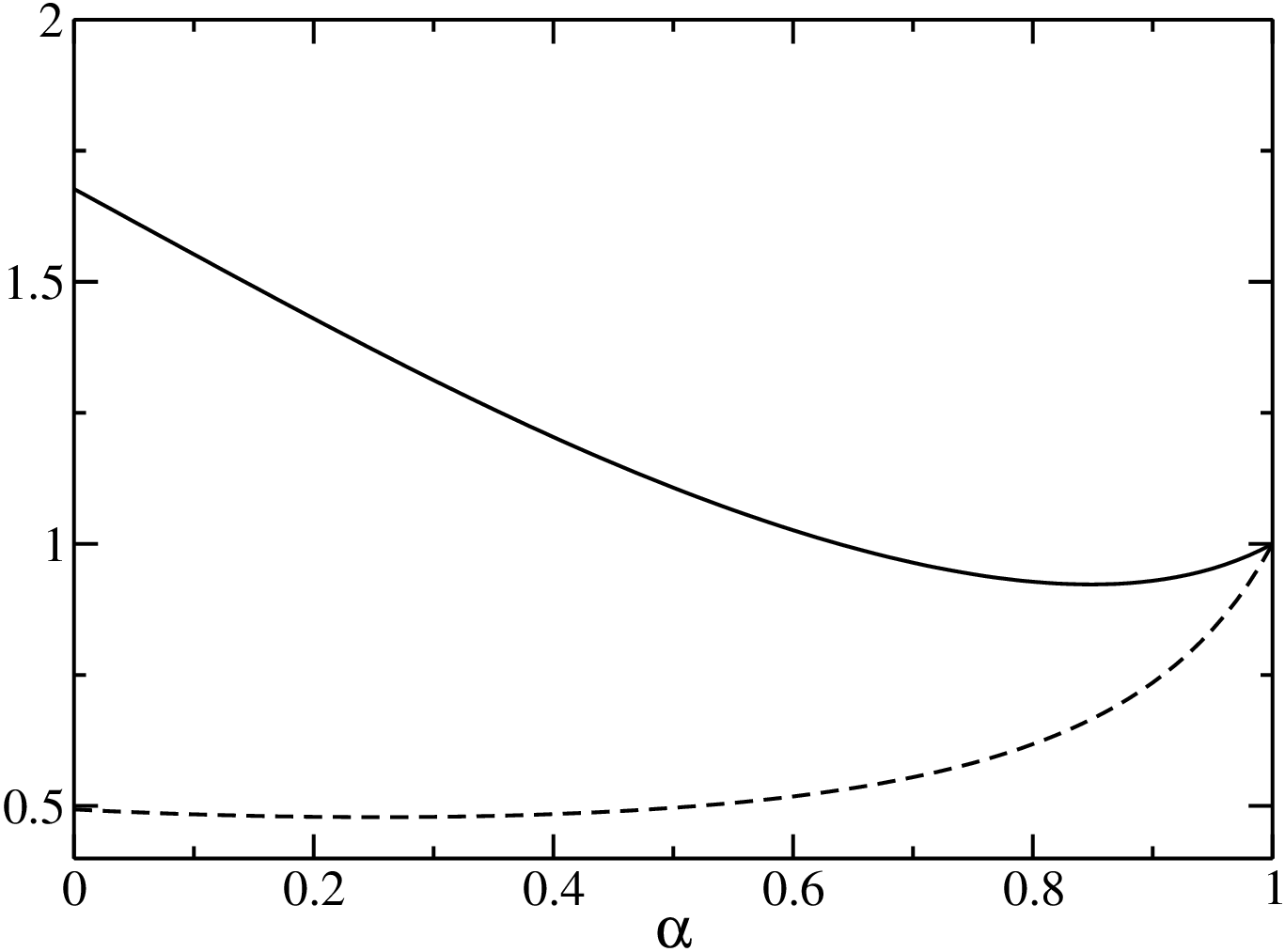}
\end{center}
\end{minipage}
\begin{minipage}{0.48\linewidth}
\begin{center}
\includegraphics[angle=0,width=0.9\linewidth]{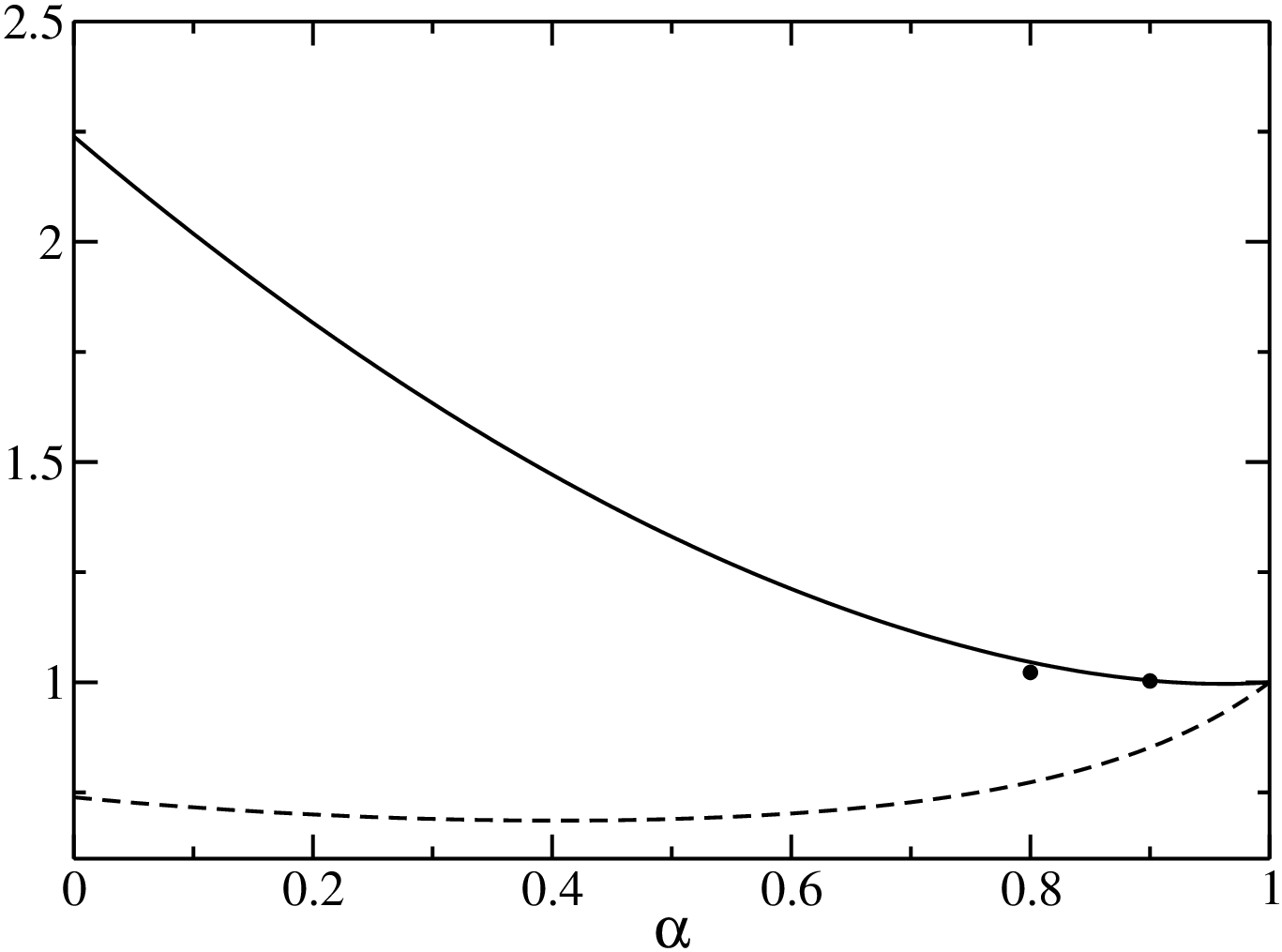}
\end{center}
\end{minipage}
\caption{Reduced conductivity of a two-dimensional system (left) and 
three-dimensional system (right) as a function of the inelasticity. The 
solid line is the theoretical prediction given by eq. (\ref{tcPrediction}), 
the dashed line is the theoretical prediction of \cite{gm02}, and the dots are 
the simulation results of \cite{vaz11}.  }\label{fig2}
\end{figure}

In Fig. \ref{fig1} the reduced viscosity $\frac{\eta(\alpha)}{\eta(1)}$ is 
plotted as a function of the inelasticity for $d=2$ and $d=3$, and in Fig. 
\ref{fig2} the same is done but for the reduced conductivity 
$\frac{\widetilde{\kappa}(\alpha)}{\widetilde{\kappa}(1)}$. 
Although strictly speaking, we do not consider the exact same
thermostating mechanism as in \cite{gm02} where the driving amplitude is
chosen to depend on the local temperature, it is nevertheless
relevant to compare our predictions to those of Ref. \cite{gm02}. 
In the case of the shear viscosity we obtain 
very similar results. The difference is due to the approximate 
method to 
evaluate the coefficient, yielding a smoother curve with the present method. 
In contrast, for the reduced conductivity, we obtain strong discrepancies.  
Equation (\ref{tcPrediction}) predicts an enhanced of the 
conductivity as the inelasticity increases while the prediction of \cite{gm02} 
goes in the opposite direction. For completeness, 
we have  reported some simulation data 
available in the literature. For the shear viscosity, the results were taken from 
Ref. \cite{gm02} for $d=3$. The agreement with the theoretical 
prediction is good. For the heat conductivity, we have taken the results of 
\cite{vaz11}, pertaining also to dissipative 
hard spheres, for the smallest density and wave vector. 
The agreement with equation (\ref{tcPrediction}) in this case is very good for 
the two values of the coefficient of normal restitution. Let us note that the 
transport coefficients were also evaluated for moderate densities in \cite{g11} 
via the Enskog equation and the density dependence agreed qualitatively well 
with the simulation results of \cite{vaz11}. 

\section{Conclusions and perspectives}\label{section5}

In this paper we have derived the Navier-Stokes hydrodynamic equations for a 
system of hard 
particles heated by the so-called stochastic thermostat. We have restricted to 
situations close to the homogeneous stationary state that the system reaches 
in the long time limit. Under these conditions, the system is described by the 
Boltzmann equation linearized around the stationary state. We could 
calculate the eigenvalues and eigenfunctions of the operator describing the 
dynamics of the system that are relevant in the hydrodynamic description. Let 
us remark that, although we are considering linear response around the 
stationary state, the modes depend on the properties of the time-dependent 
$\beta$-state through quantities related to 
$\left.\frac{\partial\chi(c,\beta)}{\partial\beta}\right|_{\beta=1}$. 
The properties of the $\beta$-state brought to the fore 
were summarized in section \ref{section2} and in particular,  
$\beta-1$ can be viewed as measuring
the distance to stationarity.
With the 
aid of these modes,
we derived the linear Navier-Stokes equations obtaining formulas for the 
transport coefficients that are expressed as Green-Kubo relations. Assuming that 
the time correlation functions that appear in these formulas decay with only 
one kinetic mode, we calculated explicitly the transport coefficients as 
functions of the inelasticity, $\alpha$, and the spatial dimension, $d$. Let us 
note that, at this level, the dynamics also depends on the properties of the 
$\beta$-state, through the transport coefficients. 
Moreover, as it is reflected in eq. (\ref{localBetaState}), the $\beta$-state 
plays the role of reference state in the sense that, in the hydrodynamic scale 
a local $\beta$-state distribution is reached. This fact is in connection with 
the results of \cite{l06} where it is seen that, close to a stationary state, 
the zeroth order in the gradients distribution is not merely the local 
stationary distribution but a more complex one (in our case played by the 
$\beta$-state). In this sense, for situations where non-linear effects are 
important, the complete hydrodynamic equations at Navier-Stokes order could 
be derived by the Chapman-Enskog scheme taking into account these ideas. We 
hope this work will contribute to stimulate more studies in this direction and 
also at the level of computer simulations by, for example, measuring the 
transport coefficients. Also, since many effects shown in the 
paper depend on the two-parameter distribution of the $\beta$-state, it is 
expected that similar phenomena will occur for other sorts of homogeneous 
thermostats. 

\section{Acknowledgments}

We would like to thank Vicente Garz\'o and 
Mois\'es Garc\'ia Chamorro for useful discussions.
This research was supported by the Ministerio de Educaci\'{o}n y
Ciencia (Spain) through Grant No. FIS2011-24460 (partially financed
by FEDER funds).

\appendix
\section{Evaluation of $\gamma$ in the first Sonine 
approximation}\label{apendiceA}
In this Appendix we calculate the coefficient $\gamma$ given by eq. 
(\ref{def.gamma}) 
\be
\gamma=3\mu(1)-\left.\frac{d\mu(\beta)}{d\beta}\right|_{\beta=1}, 
\ee
in the first Sonine approximation. By substituting the approximate expression 
of $\chi(c,\beta)$ given by eq. (\ref{sonApBeta}) into the definition of 
$\mu(\beta)$ (see eq. (\ref{def.mu})), we obtain
\be
\mu(\beta)=\mu_M+\mu_Sa_2(\beta)
\ee
where 
\be
\mu_M=-\frac{1}{2d}\int d\mathbf{c}_1\int d\mathbf{c}_2\chi_M(c_1)\chi_M(c_2)
T_0(\mathbf{c}_1,\mathbf{c}_2)(c_1^2+c_2^2), 
\ee
and 
\be
\mu_S=-\frac{1}{d}\int d\mathbf{c}_1\int d\mathbf{c}_2\chi_M(c_1)\chi_M(c_2)
S_2(c_1^2)T_0(\mathbf{c}_1,\mathbf{c}_2)(c_1^2+c_2^2). 
\ee
The integrals can be readily performed with the result
\be
\mu_M=\frac{\pi^{\frac{d-1}{2}}(1-\alpha^2)}{\sqrt{2}d\Gamma(d/2)}, \qquad
\mu_S=\frac{3\pi^{\frac{d-1}{2}}(1-\alpha^2)}{16\sqrt{2}d\Gamma(d/2)}. 
\ee
Finally, the expression for $\gamma$ is
\be
\gamma=3\mu_M+\left[3a_2(1)
-\left.\frac{da_2(\beta)}{d\beta}\right|_{\beta=1}\right]\mu_S,
\ee
where the expression of $a_2(\beta)$ has been provided in the main text.
\begin{figure}
\begin{minipage}{0.48\linewidth}
\begin{center}
\includegraphics[angle=0,width=0.9\linewidth]{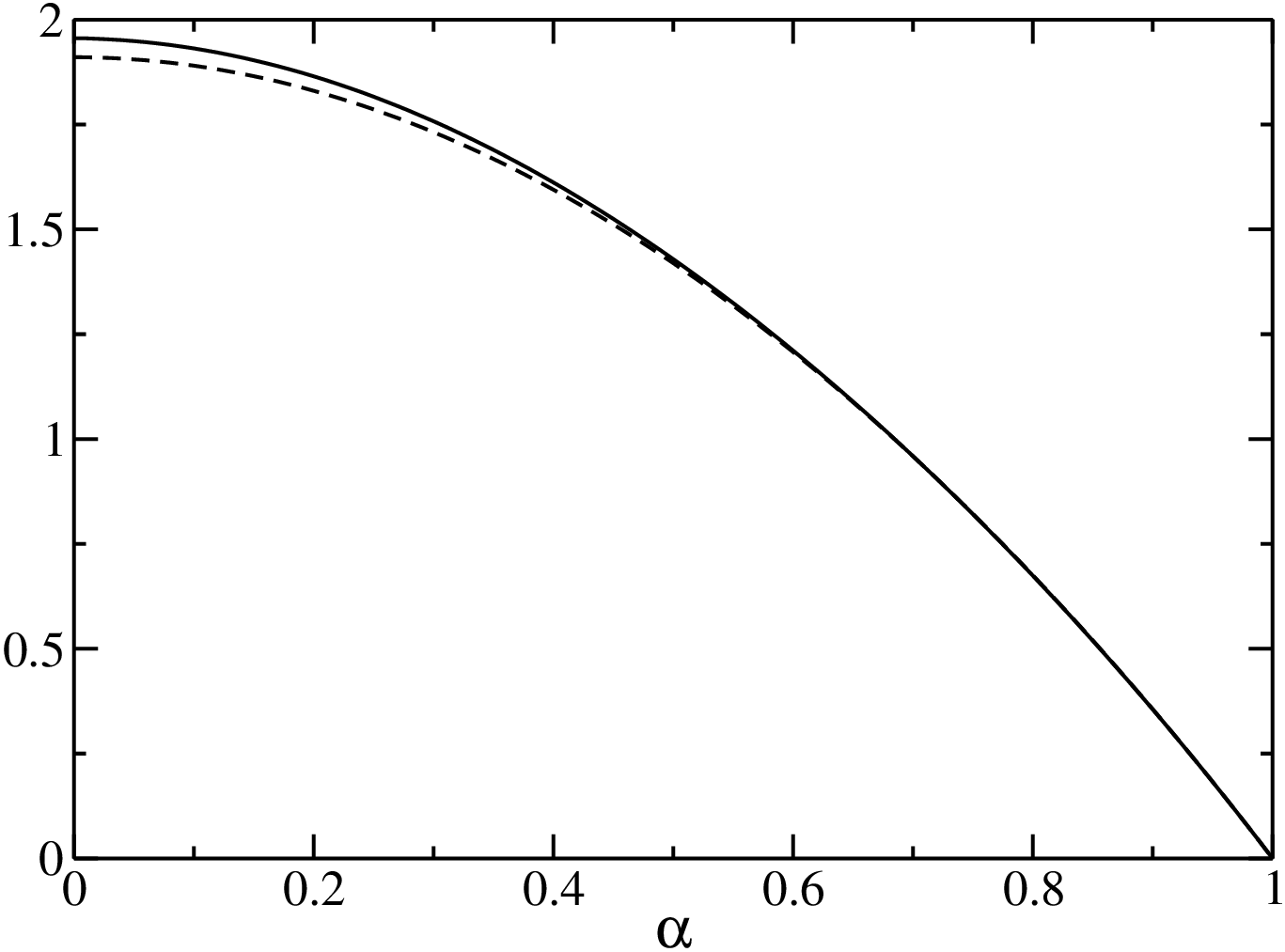}
\end{center}
\end{minipage}
\begin{minipage}{0.48\linewidth}
\begin{center}
\includegraphics[angle=0,width=0.9\linewidth]{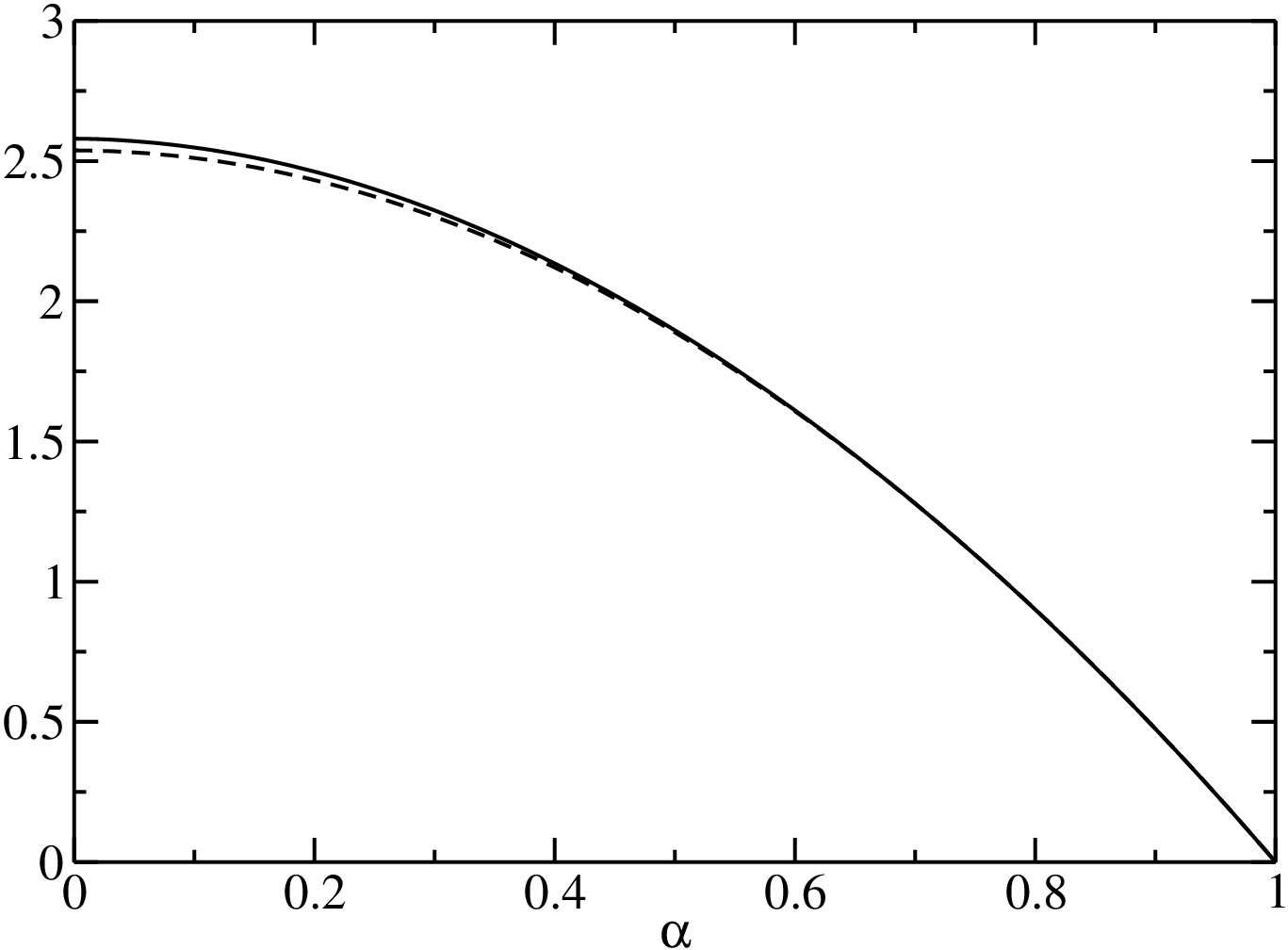}
\end{center}
\end{minipage}
\caption{Eigenvalue $\gamma$ as a function of the inelasticity for a two-dimensional 
system (left) and three-dimensional system (right). The solid line is the 
theoretical prediction for $\gamma$ and the dashed lined is the one of 
\cite{vne98}, $\gamma_{vN}$. }\label{fig3}
\end{figure}
In figure \ref{fig3}, $\gamma$ is plotted as a function of the inelasticity. The 
approximate expression of the reference \cite{vne98}, $\gamma_{vN}=3\mu(1)$, 
is also plotted, finding very similar results. Nevertheless, let us note that 
the difference between the two expressions, 
$\gamma_{vN}-\gamma=\left.\frac{da_2(\beta)}{d\beta}\right|_{\beta=1}\mu_S$ is 
of the order of $a_2(1)$.

\section{Evaluation of the hydrodynamic eigenfunctions}
\label{apendiceB}
Let us evaluate the expression
\be\label{deltafFamApp}
\delta f(\mathbf{v},t)=\frac{\bar{n}}{\bar{v}_H(t)^d}
\chi\left[\frac{\mathbf{v}}{\bar{v}_H(t)},
\frac{\bar{v}_s}{\bar{v}_H(t)}\right]-f_s(v), 
\ee
to linear order in $\delta n=\bar{n}-n$, and $\delta v_H(0)=\bar{v}_H(0)-v_s$. 
We consider the case $\mathbf{u}=\mathbf{0}$, for assuming 
$\mathbf{u}\ne\mathbf{0}$ is a straightforward generalization. Let 
us rewrite eq. (\ref{deltafFamApp}) as
\be\label{deltafFamApp2}
\delta f(\mathbf{v},t)=\frac{\bar{n}}{\bar{v}_H(t)^d}
\chi\left[\frac{\mathbf{v}}{\bar{v}_H(t)},
\frac{\bar{v}_s}{\bar{v}_H(t)}\right]
-\frac{\bar{n}}{\bar{v}_s^d}\chi_s\left(\frac{\mathbf{v}}{\bar{v}_s}\right)
+\frac{\bar{n}}{\bar{v}_s^d}\chi_s\left(\frac{\mathbf{v}}{\bar{v}_s}\right)
-f_s(v).  
\ee
Then, the first two terms on the left hand side of the equation are the 
difference between the $\beta$-state and its corresponding stationary state 
(both are linked to the same density, $\bar{n}$), while the last two terms are 
just the difference between two very close stationary states. For the last two 
terms and, taking into account that
\be\label{nvs}
nv_s^3=\bar{n}\bar{v}_s^3=\mathcal{K}(\alpha), 
\ee
as follows from from eq. (\ref{ts}) which defines the unspecified function $\mathcal{K}$, and thus
\begin{eqnarray}
\frac{\bar{n}}{\bar{v}_s^d}\chi_s\left(\frac{\mathbf{v}}{\bar{v}_s}\right)
-\frac{n}{v_s^d}\chi_s\left(\frac{\mathbf{v}}{v_s}\right)=
\frac{\mathcal{K}}{\bar{v}_s^{d+3}}
\chi_s\left(\frac{\mathbf{v}}{\bar{v}_s}\right)
-\frac{\mathcal{K}}{v_s^{d+3}}\chi_s\left(\frac{\mathbf{v}}{v_s}\right)
\nonumber\\
\approx\mathcal{K}\delta v_s\left[-\frac{(d+3)}{v_s^{d+4}}\chi_s(c)
+\frac{1}{v_s^{d+3}}\frac{\partial}{\partial v_s}
\chi_s\left(\frac{\mathbf{v}}{v_s}\right)\right], 
\end{eqnarray}
where $\delta v_s=\bar{v}_s-v_s$. It is more convenient to write this 
expression in terms of the difference in densities. Due to eq. (\ref{nvs}), 
we have
\be
\frac{\delta n}{n}=-3\frac{\delta v_s}{v_s}, 
\ee
so that
\be\label{chor1}
\frac{\bar{n}}{\bar{v}_s^d}\chi_s\left(\frac{\mathbf{v}}{\bar{v}_s}\right)
-\frac{n}{v_s^d}\chi_s\left(\frac{\mathbf{v}}{v_s}\right)\approx
\frac{n}{v_s^d}\frac{\delta n}{n}\left\{\chi_s(c)
+\frac{1}{3}\frac{\partial}{\partial\mathbf{c}}\cdot[\mathbf{c}\chi_s(c)]
\right\}. 
\ee
Now, let us evaluate the other difference
\be\label{od}
\frac{\bar{n}}{\bar{v}_H(t)^d}
\chi\left[\frac{\mathbf{v}}{\bar{v}_H(t)},
\frac{\bar{v}_s}{\bar{v}_H(t)}\right]
-\frac{\bar{n}}{\bar{v}_s^d}\chi_s\left(\frac{\mathbf{v}}{\bar{v}_s}\right)
\approx\frac{\partial}{\partial v_H(0)}\left\{\frac{n}{v_H(t)^d}
\chi\left[\frac{\mathbf{v}}{v_H(t)},\frac{v_s}{v_H(t)}\right]\right\}_s
[\bar{v}_H(0)-\bar{v}_s], 
\ee
where the subscript $s$ refers to the functional in the stationary 
state, i.e. $F[n,v_H(0)]_s=F[\bar{n},\bar{v}_s]$. The functional derivative is
\be\label{fd}
\frac{\partial}{\partial v_H(0)}\left\{\frac{n}{v_H(t)^d}
\chi\left[\frac{\mathbf{v}}{v_H(t)},\frac{v_s}{v_H(t)}\right]\right\}=
-\frac{n}{v_H(t)^{d+1}}\left\{\frac{\partial}{\partial\mathbf{c}}
\cdot[\mathbf{c}\chi(c,\beta)]
+\beta\frac{\partial}{\partial\beta}\chi(c,\beta)\right\}
\frac{\partial v_H(t)}{\partial v_H(0)}, 
\ee
while the partial derivative can be calculated taken into account eq. 
(\ref{ev.vH}), giving
\be
\left[\frac{\partial v_H(t)}{\partial v_H(0)}\right]_s
=e^{-\gamma\frac{\bar{v}_s}{\bar{\ell}}t}. 
\ee
Then, by substituting eq. (\ref{fd}) into eq. (\ref{od}) and defining 
$\widetilde{\mathbf{c}}=\mathbf{v}/\bar{v}_s$ we have 
\begin{eqnarray}\label{chor2}
\frac{\bar{n}}{\bar{v}_H(t)^d}
\chi\left[\frac{\mathbf{v}}{\bar{v}_H(t)},
\frac{\bar{v}_s}{\bar{v}_H(t)}\right]
-\frac{\bar{n}}{\bar{v}_s^d}\chi_s\left(\frac{\mathbf{v}}{\bar{v}_s}\right)
\approx-\frac{\bar{n}}{\bar{v}_s^d}
\left\{\frac{\partial}{\partial\widetilde{\mathbf{c}}}\cdot[\widetilde{\mathbf{c}}
\chi_s(\widetilde{c})]
+\left.\frac{\partial}{\partial\beta}\chi(\widetilde{c},\beta)\right|_{\beta=1}
\right\}
e^{-\gamma\frac{\bar{v}_s}{\bar{\ell}}t}
\frac{\bar{v}_H(0)-\bar{v}_s}{\bar{v}_s}\nonumber\\
\approx -\frac{n}{v_s^d}
\left\{\frac{\partial}{\partial\mathbf{c}}\cdot[\mathbf{c}\chi_s(c)]
+\left.\frac{\partial}{\partial\beta}\chi(c,\beta)\right|_{\beta=1}\right\}
e^{-\gamma\frac{v_s}{\ell}t}
\left[\frac{\delta v_H(0)}{v_s}+\frac{1}{3}\frac{\delta n}{n}\right], 
\end{eqnarray}
where we have substituted $\bar{n}$ by $n$  and $\bar{v}_s$ by 
$v_s$ (that can be done to linear order) and we have used that
\be
\frac{\bar{v}_H(0)-v_s+v_s-\bar{v}_s}{v_s}=
\frac{\delta v_H(0)}{v_s}+\frac{1}{3}\frac{\delta n}{n}. 
\ee
Taking into account (\ref{chor1}) and (\ref{chor2}), we finally arrive at the 
result reported in the main text for $\mathbf{u}=\mathbf{0}$
\begin{eqnarray}
\delta\chi(\mathbf{c},s)=\frac{\delta n}{n}\left[\chi_s(c)+\frac{1}{3}
\frac{\partial}{\partial\mathbf{c}}\cdot[\mathbf{c}\chi_s(c)]\right]
\nonumber\\
-\left[\frac{\delta n}{3n}+\frac{\delta v_H(0)}{v_s}\right]e^{-\gamma s}
\left[\frac{\partial}{\partial\mathbf{c}}\cdot[\mathbf{c}\chi_s(c)]
+\left.\frac{\partial}{\partial\beta}\chi(c,\beta)\right|_{\beta=1}\right]. 
\end{eqnarray} 

\section{Evaluation of the fluxes to first order in $k$}
\label{apendiceC}

In this appendix, we evaluate the fluxes given by  eq. (\ref{fluxes}) to first 
order in $k$. The function $\mathcal{Q}\delta\chi_{\mathbf{k}}$ fulfills eq. 
(\ref{ec.Qdchi}) and can be integrated formally as 
\be
\mathcal{Q}\delta\chi_{\mathbf{k}}(\mathbf{c},s)=
e^{\mathcal{Q}(\Lambda-i\mathbf{k}\cdot\mathbf{c})s}
\mathcal{Q}\delta\chi_{\mathbf{k}}(\mathbf{c},0)
-\int_0^sds'e^{\mathcal{Q}(\Lambda-i\mathbf{k}\cdot\mathbf{c})\mathcal{Q}(s-s')}
\mathcal{Q}i\mathbf{k}\cdot\mathbf{c}\mathcal{P}
\delta\chi_{\mathbf{k}}(\mathbf{c},s'). 
\ee
Choosing the initial condition in the hydrodynamic subspace, i.e. 
$\mathcal{Q}\delta\chi_{\mathbf{k}}(\mathbf{c},0)=0$, and neglecting the $k$ 
contribution in the kinetic modes, we have
\be\label{Qdeltachi1App}
\mathcal{Q}\delta\chi_{\mathbf{k}}(\mathbf{c},s)\approx
-\int_0^sds'e^{\mathcal{Q}\Lambda\mathcal{Q}(s-s')}
\mathcal{Q}i\mathbf{k}\cdot\mathbf{c}\mathcal{P}
\delta\chi_{\mathbf{k}}(\mathbf{c},s'). 
\ee
Inserting the above equation into the expression of the pressure tensor and 
taking into account that
\be
\int d\mathbf{c}\Delta_{jp}(\mathbf{c})\mathcal{Q}\Lambda(\mathbf{c})\mathcal{Q}
h(\mathbf{c})
=\int d\mathbf{c}\Delta_{jp}(\mathbf{c})\Lambda(\mathbf{c})h(\mathbf{c}), 
\ee 
we obtain that
\begin{eqnarray}
\Pi_{\mathbf{k}, jp}(s)\approx -\int d\mathbf{c}\Delta_{jp}(\mathbf{c})
\int_0^sds'e^{\Lambda(s-s')}i\mathbf{k}\cdot\mathbf{c}
\mathcal{P}\delta\chi_{\mathbf{k}}(\mathbf{c},s')\nonumber\\
=-\int d\mathbf{c}\Delta_{jp}(\mathbf{c})
\int_0^sds'e^{\Lambda(s-s')}i\mathbf{k}\cdot\mathbf{c}\sum_{\beta=1}^{d+2}
\langle\bar{\xi}_{\beta}(\mathbf{c})|\delta\chi_{\mathbf{k}}(\mathbf{c},s')\rangle
\xi_{\beta}(\mathbf{c})\nonumber\\
=-\sum_r\sum_qik_r\int_0^sds'w_{\mathbf{k},q}(s')\int d\mathbf{c}
\Delta_{jp}(\mathbf{c})e^{\Lambda(s-s')}c_r\xi_{2,q}(\mathbf{c}),
\end{eqnarray}
where we have taken into account that there is no coupling with the density nor
with the temperature. Finally, by symmetry considerations, we arrive at the 
expression used in the main text
\be
\Pi_{\mathbf{k}, jp}(s)\approx -i\int_0^sds'G_{xy}(s-s')
[k_jw_{\mathbf{k},p}(s')+k_pw_{\mathbf{k},j}(s')
-\frac{2}{d}\delta_{jp}\mathbf{k}\cdot\mathbf{w}_{\mathbf{k}}(s')], 
\ee
where
\be
G_{xy}(s)=\int d\mathbf{c}\Delta_{xy}(\mathbf{c})e^{\Lambda s}c_x
\xi_{2,y}(\mathbf{c}). 
\ee

To evaluate the heat flux to first order in $k$, we substitute expression 
(\ref{Qdeltachi1App}) into the heat flux and taking into account 
\be
\int d\mathbf{c}\Sigma_{j}(\mathbf{c})\mathcal{Q}\Lambda(\mathbf{c})\mathcal{Q}
h(\mathbf{c})
=\int d\mathbf{c}\Sigma_{j}(\mathbf{c})\Lambda(\mathbf{c})h(\mathbf{c}), 
\ee 
we obtain
\begin{eqnarray}
&&\phi_{\mathbf{k}, j}(s)\approx -\int d\mathbf{c}\Sigma_{j}(\mathbf{c})
\int_0^sds'e^{\Lambda(s-s')}i\mathbf{k}\cdot\mathbf{c}
\mathcal{P}\delta\chi_{\mathbf{k}}(\mathbf{c},s')\nonumber\\
&&=-\int d\mathbf{c}\Sigma_{j}(\mathbf{c})
\int_0^sds'e^{\Lambda(s-s')}i\mathbf{k}\cdot\mathbf{c}\sum_{\beta=1}^{d+2}
\langle\bar{\xi}_{\beta}(\mathbf{c})|\delta\chi_{\mathbf{k}}(\mathbf{c},s')\rangle
\xi_{\beta}(\mathbf{c})\nonumber\\
&&=-ik_j\int d\mathbf{c}\Sigma_j(\mathbf{c})\int_0^sds'e^{\Lambda(s-s')}c_j
[\langle\bar{\xi}_1(\mathbf{c})|\delta\chi_{\mathbf{k}}(\mathbf{c},s')\rangle
\xi_1(c)+
\langle\bar{\xi}_3(\mathbf{c})|\delta\chi_{\mathbf{k}}(\mathbf{c},s')\rangle
\xi_3(c)], \nonumber\\
\end{eqnarray}
where we have used that there is no coupling with the flow velocity. Finally, 
taking into account eqs. (\ref{comp1}) and (\ref{comp3}) and symmetry 
considerations, we arrive at the expression of the main text
\be
\phi_{\mathbf{k},j}(s)=-ik_j\int_0^sds'
\left\{\rho_{\mathbf{k}}(s')\left[H_1(s-s')+\frac{1}{3}H_3(s-s')\right]
+\frac{1}{2}\theta_{\mathbf{k}}(s')H_3(s-s')\right\}, 
\ee
where
\be
H_j(s)=\int d\mathbf{c}\Sigma_{x}(\mathbf{c})e^{\Lambda s}
c_x\xi_j(\mathbf{c}), \qquad j=1,3.
\ee

\section{Evaluation of the cooling rate to second order in $k$}
\label{apendiceD}

We work out here the cooling rate given by  eq. 
(\ref{coolingRate.def}) to second order in $k$. As in Appendix \ref{apendiceC}, 
choosing an initial condition with 
$\mathcal{Q}\delta\chi_{\mathbf{k}}(\mathbf{c},0)=0$, and neglecting the $k$ 
contribution in the kinetic modes,  we 
have to $k^2$ order in the hydrodynamic fields
\begin{eqnarray}\label{deltachi2}
&&\mathcal{Q}\delta\chi_{\mathbf{k}}(\mathbf{c},s)\approx
-\int_0^sds'e^{\mathcal{Q}\Lambda\mathcal{Q}(s-s')}
\mathcal{Q}i\mathbf{k}\cdot\mathbf{c}\mathcal{P}
\delta\chi_{\mathbf{k}}(\mathbf{c},s')\nonumber\\
&&-\int_0^sds'\int_0^{s-s'}ds''\mathcal{Q}e^{\Lambda(s-s'-s'')}
\mathbf{k}\cdot\mathbf{c}\mathcal{Q}e^{\Lambda s''}
\mathbf{k}\cdot\mathbf{c}\mathcal{P}
\delta\chi_{\mathbf{k}}(\mathbf{c},s'), 
\end{eqnarray}
where the identity for any two operators $A$ and $B$
\be
e^{(A+B)t}=e^{At}+\int_0^tdt'e^{A(t-t')}Be^{(A+B)t'}, 
\ee
has been used in order to perform the expansion. The first term of the right 
hand side of (\ref{deltachi2}) is the first order in $k$ contribution, 
$\mathcal{Q}\delta\chi_{\mathbf{k}}(\mathbf{c},s)^{(1)}$, 
calculated in Appendix \ref{apendiceC}, while the second term is the $k^2$ 
contribution, $\mathcal{Q}\delta\chi_{\mathbf{k}}(\mathbf{c},s)^{(2)}$. By 
substituting (\ref{deltachi2}) into (\ref{coolingRate.def}) the cooling rate 
to $k^2$ order is obtained. The first order contribution is
\begin{eqnarray}
\delta\zeta_{\mathbf{k}}^{(1)}(s)=-\int d\mathbf{c}\frac{2c^2}{d}
\Lambda(\mathbf{c})\int_0^sds'\mathcal{Q}e^{\Lambda(s-s')}
i\mathbf{k}\cdot\mathbf{c}\mathcal{P}\delta\chi_{\mathbf{k}}(\mathbf{c},s')
\nonumber\\
=-2\langle\bar{\xi}_3(c)|\Lambda(\mathbf{c})\int_0^sds'\mathcal{Q}e^{\Lambda(s-s')}
i\mathbf{k}\cdot\mathbf{c}\mathcal{P}\delta\chi_{\mathbf{k}}(\mathbf{c},s')
\rangle, 
\end{eqnarray}
where we have used eq. (\ref{barxidef}). 
Taking into account that
\be
\langle\bar{\xi}_3(c)|\Lambda(\mathbf{c})\mathcal{Q}h(\mathbf{c})\rangle=
\langle\bar{\xi}_3(c)|[\Lambda(\mathbf{c})-\lambda_3]h(\mathbf{c})\rangle, 
\ee
it can be rewritten as
\begin{eqnarray}\label{deltazeta1}
\delta\zeta_{\mathbf{k}}^{(1)}(s)=-2\langle\bar{\xi}_3(c)|[\Lambda(\mathbf{c})
-\lambda_3]\int_0^sds'e^{\Lambda(s-s')}i\mathbf{k}\cdot\mathbf{c}
\mathcal{P}\delta\chi_{\mathbf{k}}(\mathbf{c},s')\rangle\nonumber\\
=-2i\sum_pk_p\int_0^sds'w_{\mathbf{k},p}(s')\langle\bar{\xi}_3(c)|
[\Lambda(\mathbf{c})-\lambda_3]e^{\Lambda(s-s')}c_p\xi_{2,p}(\mathbf{c})\rangle, 
\end{eqnarray}
where we have used that, by symmetry considerations, there is only coupling 
with the flow velocity. Finally, we can transform (\ref{deltazeta1}) to write 
it as in the main text
\be
\delta\zeta_{\mathbf{k}}(s)^{(1)}=-2i\int_0^sds'
\mathbf{k}\cdot\mathbf{w}_{\mathbf{k}}(s')Z(s-s'), 
\ee
where 
\be
Z(s)=\langle\bar{\xi}_3(\mathbf{c})|[\Lambda(\mathbf{c})-\lambda_3]
e^{\Lambda s}c_x\xi_{2,x}(\mathbf{c})\rangle. 
\ee

By similar manipulations, the second order in $k$ contribution can be written 
as 
\begin{eqnarray}\label{deltazeta2}
&&\delta\zeta_{\mathbf{k}}^{(2)}(s)=-2\langle\bar{\xi}_3(c)|[\Lambda(\mathbf{c})
-\lambda_3]\int_0^sds'\int_0^{s-s'}ds''e^{\Lambda(s-s'-s'')}\mathbf{k}\cdot\mathbf{c}
\mathcal{Q}e^{\Lambda s''}  \mathbf{k}\cdot\mathbf{c}
\mathcal{P}\delta\chi_{\mathbf{k}}(\mathbf{c},s')\rangle\nonumber\\
&&=-2\langle\bar{\xi}_3(c)|[\Lambda(\mathbf{c})-\lambda_3]\int_0^sds'
\int_0^{s-s'}ds''e^{\Lambda(s-s'-s'')}\mathbf{k}\cdot\mathbf{c}
\mathcal{Q}e^{\Lambda s''}  \mathbf{k}\cdot\mathbf{c}
[\rho_{\mathbf{k}}(s')\xi_1(c)
+\langle\bar{\xi}_3(c)|\delta\chi_{\mathbf{k}}(\mathbf{c},s')\rangle\xi_3(c)]
\rangle\nonumber\\
&&=-2k^2\langle\bar{\xi}_3(c)|[\Lambda(\mathbf{c})-\lambda_3]\int_0^sds'
\int_0^{s-s'}ds''e^{\Lambda(s-s'-s'')}c_x
\mathcal{Q}e^{\Lambda s''}c_x
[\rho_{\mathbf{k}}(s')\xi_1(c)
+\langle\bar{\xi}_3(c)|\delta\chi_{\mathbf{k}}(\mathbf{c},s')\rangle\xi_3(c)]
\rangle. \nonumber\\
\end{eqnarray}
that can be recast as 
\be
\delta\zeta_{\mathbf{k}}^{(2)}(s)=-2k^2\int_0^sds'
[\rho_{\mathbf{k}}(s')Z_1(s-s')+
\langle\bar{\xi}_3(c)|\delta\chi_{\mathbf{k}}(\mathbf{c},s')\rangle Z_3(s-s')], 
\ee
where 
\be
Z_j(s)=\langle\bar{\xi}_3(\mathbf{c})|[\Lambda(\mathbf{c})-\lambda_3]\int_0^s
ds'e^{\Lambda (s-s')}c_x\mathcal{Q}e^{\Lambda s'}
c_x\xi_{j}(\mathbf{c})\rangle, \quad j=1,3.  
\ee
or, as written in the main text 
\be
\delta\zeta_{\mathbf{k}}^{(2)}(s)=-2k^2\int_0^sds'
\left\{\rho_{\mathbf{k}}(s')\left[Z_1(s-s')+\frac{1}{3}Z_3(s-s')\right]
+\frac{1}{2}\theta_{\mathbf{k}}(s')Z_3(s-s')\right\}. 
\ee

\section{Analysis of the equation for the transversal velocity}
\label{apendiceE}

In Laplace space, the transversal velocity is given by eq. (\ref{eHLw})
\be\label{wtras}
\bar{w}_{\mathbf{k},\perp}(z)=\frac{w_{\mathbf{k},\perp}(0)}
{z+k^2\bar{G}_{xy}(z)}, 
\ee
where we have skipped the superscript $j$. This formula depends on 
$\bar{G}_{xy}(z)$ that, in principle, can be a complicated object. We 
already know that in real time it decays with the kinetic modes and in the 
free cooling case it has been shown numerically that it is very well fitted by 
a single exponential (single mode approximation). Let us consider this simple 
form for the general case can be performed in a similar way. In this 
approximation, we have
\be
G_{xy}(s)=Ce^{-\lambda s}, \qquad \bar{G}_{xy}(z)=\frac{C}{z+\lambda}, 
\ee
and eq. (\ref{wtras}) is just
\be\label{wtrassing}
\bar{w}_{\mathbf{k},\perp}(z)=\frac{w_{\mathbf{k},\perp}(0)(z+\lambda)}
{z(z+\lambda)+Ck^2}, 
\ee
that can be inverted exactly in terms of two exponentials 
\cite{HansenMcDonald}. The function given by eq. (\ref{wtrassing}) has the two 
poles
\begin{eqnarray}
z_1(k)&=&\frac{-\lambda+\sqrt{\lambda^2-4Ck^2}}{2}\approx-\frac{C}{\lambda}k^2, 
\\
z_2(k)&=&\frac{-\lambda-\sqrt{\lambda^2-4Ck^2}}{2}\approx-\lambda
+\frac{C}{\lambda}k^2, 
\end{eqnarray}
and it can be written as
\be\label{wdes}
\bar{w}_{\mathbf{k},\perp}(z)=\frac{A(k)}{z-z_1(k)}
+\frac{B(k)}{z-z_2(k)}. 
\ee
Then, we have
\begin{eqnarray}
&&w_{\mathbf{k},\perp}(0)(z+\lambda)=A(k)[z-z_2(k)]+B(k)[z-z_1(k)]\nonumber\\
&&\approx
A(k)\left(z+\lambda-\frac{C}{\lambda}k^2\right)
+B(k)\left(z+\frac{C}{\lambda}k^2\right), 
\end{eqnarray}
with which we identify the constant $A$ and $B$ to zeroth order in $k$
\be\label{const}
A\approx w_{\mathbf{k},\perp}(0), \qquad B\approx 0. 
\ee
With eqs. (\ref{wdes}) and (\ref{const}) we obtain the expression for the 
transversal velocity of the main text, eq. (\ref{vtransInt}), 
\be\label{vtransIntApp}
w_{\mathbf{k},\perp}(s)\approx 
w_{\mathbf{k},\perp}(0)e^{-\eta k^2s}, 
\ee
with
\be
\eta=\frac{C}{\lambda}=\int_0^\infty dsG_{xy}(s). 
\ee

If the function $G_{xy}(s)$ is a linear combination of kinetic modes, the 
analysis can be performed following the same lines obtaining eq. 
(\ref{vtransIntApp}) with $\eta=\int_0^\infty dsG_{xy}(s)$. 

\section{Analysis of the coupled hydrodynamic equations}
\label{apendiceF}

We evaluate here the asymptotic behavior of eqs. (\ref{eHLcouple})
\be\label{eHLcoupleappen}
[zI+A(k,z)]\left(\begin{array}{c}
\bar{\rho}_{\mathbf{k}}(z)\\
\bar{w}_{\mathbf{k}, ||}(z)\\
\bar{\theta}_{\mathbf{k}}(z)\end{array}\right)=
\left(\begin{array}{c}\rho_{\mathbf{k}}(0)\\w_{\mathbf{k}, ||}(0)\\
\theta_{\mathbf{k}}(0)\end{array}\right), 
\ee
in the hydrodynamic limit. The matrix $A(k,z)$ is given in eq. (\ref{matrixA}) 
and can be written as
\be\label{matrixExpan}
A(k,z)=A_0+ikA_1(z)+k^2A_2(z), 
\ee
where
\be
A_0=\left(\begin{array}{ccc} 0 & 0 & 0 \\
0 & 0 & 0 \\
\frac{2}{3}\gamma & 0 & \gamma\end{array}\right), 
\qquad
A_1(z)=\left(\begin{array}{ccc} 0 & 1 & 0 \\
\frac{1}{2} & 0 & \frac{1}{2} \\
0 & \bar{q}(z) & 0\end{array}\right),
\ee
\be
A_2(z)=\left(\begin{array}{ccc} 0 & 0 & 0 \\
0 & 2\frac{d-1}{d}\bar{G}_{xy}(z) & 0 \\
\frac{2}{d}\bar{G}_1(z) & 0  & \frac{2}{d}\bar{G}_3(z)\end{array}\right). 
\ee

In terms of the eigenvalues and eigenfunctions of 
the matrix $A(k,z)$
\be
A(k,z)|\psi_{\beta}(k,z)\rangle=a_{\beta}(k,z)|\psi_{\beta}(k,z)\rangle, 
\ee
the solution can be written explicitly as
\be
|\bar{y}(\mathbf{k},z)\rangle=[zI+A(k,z)]^{-1}|y(\mathbf{k},0)\rangle
=\sum_{\beta=1}^3
\frac{\langle\bar{\psi}_{\beta}(k,z)|y(\mathbf{k},0)\rangle}{z+a_{\beta}(k,z)}
|\psi_{\beta}(k,z)\rangle, 
\ee
where we have introduced the notation
\be
|y(\mathbf{k},s)\rangle\equiv\left(\begin{array}{c}
\rho_{\mathbf{k}}(s)\\
w_{\mathbf{k}, ||}(s)\\
\theta_{\mathbf{k}}(s)\end{array}\right), \qquad
|\bar{y}(\mathbf{k},z)\rangle\equiv\left(\begin{array}{c}
\bar{\rho}_{\mathbf{k}}(z)\\
\bar{w}_{\mathbf{k}, ||}(z)\\
\bar{\theta}_{\mathbf{k}}(z)\end{array}\right),  
\ee
and the functions $\{\langle\bar{\psi}_{\beta}(k,z)|\}_{\beta=1}^3$ are the left 
eigenfunctions of $A(k,z)$. Let us introduce the expansion in powers of $k$ 
of the eigenvalues and eigenfunctions
\begin{eqnarray}
|\psi_{\beta}(k,z)\rangle&=&|\psi_{\beta}^{(0)}(z)\rangle
+k|\psi_{\beta}^{(1)}(z)\rangle+k^2|\psi_{\beta}^{(2)}(z)\rangle+\dots, \\
a_{\beta}(k,z)&=&a_{\beta}^{(0)}(z)+ka_{\beta}^{(1)}(z)+k^2a_{\beta}^{(2)}(z)
+\dots. 
\end{eqnarray}
In the hydrodynamic limit, and applying the same ideas as in Appendix 
\ref{apendiceE}, we obtain 
\be
|\bar{y}(\mathbf{k},z)\rangle\approx\sum_{\beta}
\frac{\langle\phi_{\beta}|y(\mathbf{k},0)\rangle}
{z-\lambda_{\beta}(k)}|\psi_{\beta}^{(0)}\rangle, 
\ee
where $\{\lambda_{\beta}(k)\}_{\beta=1}^3$ are the hydrodynamic eigenvalues that 
appear as the smallest root of eq. $z+a_{\beta}(k,z)=0$. The set 
$\{\langle\phi_{\beta}|\}_{\beta=1}^3$ is the bi-orthogonal set constructed 
to have $\langle\phi_{\beta}|\psi_{\beta'}^{(0)}\rangle=\delta_{\beta,\beta'}$. The 
problem is then to calculate the sets 
$\{|\psi_{\beta}^{(0)}\rangle\}_{\beta=1}^3$, $\{\langle\phi_{\beta}|\}_{\beta=1}^3$ and 
$\{a_{\beta}(k,z)\}_{\beta=1}^3$. As $\{\lambda_{\beta}(k)\}_{\beta=1}^3$ is needed 
to $k^2$ order, $\{a_{\beta}(k,z)\}_{\beta=1}^3$ have to be calculated at the same 
order, which can be done by standard perturbation theory. 

The eigenvalues of $A_0$ can be easily calculated, obtaining
\be
a_1^{(0)}=a_2^{(0)}=0, \qquad a_3^{(0)}=\gamma, 
\ee
so that the vanishing eigenvalue is two-fold degenerate. The corresponding 
eigenfunctions are
\be
|u_1\rangle=\left(\begin{array}{c}
3 \\ 0 \\ -2 \end{array}\right), \qquad
|u_2\rangle=\left(\begin{array}{c}
3 \\ 1 \\ -2 \end{array}\right), \qquad
|u_3\rangle=|\psi_3^{(0)}\rangle=\left(\begin{array}{c}
0 \\ 0 \\ 1 \end{array}\right),  
\ee 
and the bi-orthogonal set is
\be
\langle v_1|=\left(\frac{1}{3}, -1, 0 \right), \qquad
\langle v_2|=\left(0, 1, 0 \right), \qquad
\langle v_3|=\langle\phi_3|=\left(\frac{2}{3}, 0, 1 \right). 
\ee
In the non-degenerate case, the first contributions to the expansion are
\be
a_3^{(1)}(z)=\langle v_3|iA_1|u_3\rangle=0, 
\ee
and 
\be
a_3^{(2)}(z)=\langle v_3|A_2|u_3\rangle
+\frac{1}{\gamma}
\sum_{n\ne 3}\langle v_n|iA_1|u_3\rangle\langle v_3|iA_1|u_n\rangle
=\frac{2}{d}\bar{G}_3(z)-\frac{1}{3\gamma}-\frac{q(z)}{2\gamma}, 
\ee
so that we have
\be\label{a3approx}
a_3(k,z)\approx\gamma+\left[\frac{2}{d}\bar{G}_3(z)-\frac{1}{3\gamma}
-\frac{q(z)}{2\gamma}\right]k^2. 
\ee
For the degenerate case, we first have to consider the sub-matrix
\be
iA_1^{(S)}(z)=\left(\begin{array}{cc} 
\langle v_1|iA_1|u_1\rangle & \langle v_1|iA_1|u_2\rangle \\
\langle v_2|iA_1|u_1\rangle & \langle v_2|iA_1|u_2\rangle  \\
\end{array}\right)
=i\left(\begin{array}{cc} 
-\frac{1}{2} & -\frac{1}{6} \\
\frac{1}{2} & \frac{1}{2} \\
\end{array}\right). 
\ee
Its eigenvalues are the first order correction of the degenerate eigenvalues
\be
a_1^{(1)}(z)=-\frac{i}{\sqrt{6}}, \qquad a_2^{(1)}(z)=\frac{i}{\sqrt{6}},
\ee
and its eigenfunctions give us the components of the corresponding zeroth 
order eigenfunction
\be
\left(\begin{array}{c}
\langle v_1|\psi_1^{(0)}\rangle \\ \langle v_2|\psi_1^{(0)}\rangle 
\end{array}\right)
=\left(\begin{array}{c}
-1-\sqrt{\frac{2}{3}}  \\  1 \end{array}\right), \qquad
 \left(\begin{array}{c}
\langle v_1|\psi_2^{(0)}\rangle \\ \langle v_2|\psi_2^{(0)}\rangle 
\end{array}\right)
=\left(\begin{array}{c}
-1+\sqrt{\frac{2}{3}} \\ 1 \end{array}\right), 
\ee 
so that 
\be
|\psi_1^{(0)}\rangle=\left(\begin{array}{c}
-6 \\ \sqrt{6} \\ 4 \end{array}\right), \qquad
|\psi_2^{(0)}\rangle=\left(\begin{array}{c}
6 \\ \sqrt{6} \\ -4 \end{array}\right), 
\ee
that do not depend on $z$. The corresponding bi-orthogonal functions can also be 
calculated obtaining 
\be
\langle\phi_1|=\left(-\frac{1}{12}, \frac{1}{2\sqrt{6}}, 0 \right), \qquad
\langle\phi_2|=\left(\frac{1}{12}, \frac{1}{2\sqrt{6}}, 0 \right). 
\ee
With these functions, the $k^2$ corrections to the degenerate eigenvalues follow 
from straightforward calculations
\be
a_1^{(2)}(z)=\langle\phi_1|A_2|\psi_1^{(0)}\rangle-\frac{1}{\gamma}
\langle\phi_3|iA_1|\psi_1^{(0)}\rangle
\langle\phi_1|iA_1|\psi_3^{(0)}\rangle
=\frac{d-1}{d}\bar{G}_{xy}(z)+\frac{1}{4\gamma}\left[\frac{2}{3}+q(z)\right], 
\ee
and 
\be
a_2^{(2)}(z)=\langle\phi_2|A_2|\psi_2^{(0)}\rangle-\frac{1}{\gamma}
\langle\phi_3|iA_1|\psi_2^{(0)}\rangle
\langle\phi_2|iA_1|\psi_3^{(0)}\rangle
=\frac{d-1}{d}\bar{G}_{xy}(z)+\frac{1}{4\gamma}\left[\frac{2}{3}+q(z)\right], 
\ee
so that we have
\begin{eqnarray}\label{a1approx}
a_1(k,z)&\approx&-\frac{i}{\sqrt{6}}k+\left[\frac{d-1}{d}\bar{G}_{xy}(z)
+\frac{1}{6\gamma}+\frac{q(z)}{4\gamma}\right]k^2, \\
\label{a2approx}
a_2(k,z)&\approx&+\frac{i}{\sqrt{6}}k+\left[\frac{d-1}{d}\bar{G}_{xy}(z)
+\frac{1}{6\gamma}+\frac{q(z)}{4\gamma}\right]k^2. 
\end{eqnarray}

The smallest root of  $z+a_{\beta}(k,z)=0$,
with $a_{\beta}(k,z)$ given by eqs. (\ref{a1approx}), (\ref{a2approx}) and 
(\ref{a3approx}) are the hydrodynamic eigenvalues given in the main text, 
i.e. eqs. (\ref{lambda1approx})-(\ref{lambda3approx}).

\section{Evaluation of the kinetic eigenvalues}
\label{apendiceG}

Let us assume
\be\label{aprox2appen}
\Lambda^{+}(\mathbf{c})\chi_s(c)\Delta_{jp}(\mathbf{c})\approx
\lambda_{NH}^{(1)}\chi_s(c)\Delta_{jp}(\mathbf{c}). 
\ee
Multiplying by $c_xc_y$ and integrating we obtain
\be
\lambda_{NH}^{(1)}\int d\mathbf{c}c_x^2c_y^2\chi_s(c)
=\int d\mathbf{c}c_xc_y\Lambda^{+}(\mathbf{c})\chi_s(c) c_xc_y
=\int d\mathbf{c}c_xc_y\Lambda(\mathbf{c})\chi_s(c) c_xc_y, 
\ee
so that we have
\be
\lambda_{NH}^{(1)}=\frac{4}{1+a_2^s}I_1, \qquad
I_1=\int d\mathbf{c}c_xc_y\Lambda(\mathbf{c})\chi_s(c) c_xc_y. 
\ee
The heating does not contribute to the integral $I_1$ and we have
\be
I_1
=\int d\mathbf{c}_1\int d\mathbf{c}_2c_{1x}c_{1y}
\bar{T}_0(\mathbf{c}_1,\mathbf{c}_2)(1+P_{12})\chi_s(c_1)\chi_s(c_2)
c_{1x}c_{1y}. 
\ee
Taking into account 
\be
(b_{\sigma}-1)(c_{1x}c_{1y}+c_{2x}c_{2y})=
\frac{(1+\alpha)^2}{2}(\hat{\boldsymbol{\sigma}}\cdot\mathbf{c}_{12})^2
\hat{\sigma}_x\hat{\sigma}_y
-\frac{1+\alpha}{2}(\hat{\boldsymbol{\sigma}}\cdot\mathbf{c}_{12})
(c_{12y}\hat{\sigma}_x+c_{12x}\hat{\sigma}_y), 
\ee
and the solid angle integrals
\begin{eqnarray}
\int d\hat{\boldsymbol{\sigma}}
\Theta(\hat{\boldsymbol{\sigma}}\cdot\mathbf{c}_{12})
(\hat{\boldsymbol{\sigma}}\cdot\mathbf{c}_{12})^3\hat{\sigma}_x\hat{\sigma}_y
&=&\frac{3\pi^{\frac{d-1}{2}}}{2\Gamma\left(\frac{d+5}{2}\right)}
c_{12}c_{12x}c_{12y}, \\
\int d\hat{\boldsymbol{\sigma}}
\Theta(\hat{\boldsymbol{\sigma}}\cdot\mathbf{c}_{12})
(\hat{\boldsymbol{\sigma}}\cdot\mathbf{c}_{12})^2\hat{\sigma}_j
&=&\frac{\pi^{\frac{d-1}{2}}}{\Gamma\left(\frac{d+3}{2}\right)}
c_{12}c_{12j}, 
\end{eqnarray}
we finally obtain
\begin{eqnarray}
&&I_1=\left[\frac{3(1+\alpha)^2}{4\Gamma\left(\frac{d+5}{2}\right)}
-\frac{1+\alpha}{\Gamma\left(\frac{d+3}{2}\right)}\right]\pi^{\frac{d-1}{2}}
\int d\mathbf{c}_1\int d\mathbf{c}_2\chi_s(c_1)\chi_s(c_2)c_{1x}c_{1y}c_{12}
c_{12x}c_{12y}\nonumber\\
&&=-\frac{(2d+3-3\alpha)(1+\alpha)\pi^{\frac{d-1}{2}}}{2\sqrt{2}d(d+2)\Gamma(d/2)}
\left(1+\frac{23}{16}a_2^s\right), 
\end{eqnarray}
where the last integral has been performed with $\chi_s(c)$ in the first Sonine 
approximation. 

To calculate $\lambda_{NH}^{(2)}$ we multiply eq. 
\be\label{aprox2appen2}
\Lambda^{+}(\mathbf{c})\chi_s(c)\Sigma_{j}(\mathbf{c})\approx
\lambda_{NH}^{(2)}\chi_s(c)\Sigma_{j}(\mathbf{c}), 
\ee
by $c_x$, and proceed with an integration over $\mathbf{c}$ to obtain
\be
\lambda_{NH}^{(2)}=\frac{4}{(d+2)a_2^s}I_2, \qquad
I_2=\int d\mathbf{c}c_xc^2\Lambda(\mathbf{c})\chi_s(c) c_x. 
\ee
The heating contribution to $I_2$ is simply
\be
I_2^{(H)}=\int d\mathbf{c}c_xc^2\frac{\widetilde{\xi^2}}{2}
\frac{\partial^2}{\partial c^2}\chi_s(c)c_x=\frac{d+2}{2}\widetilde{\xi^2}, 
\ee
where $\widetilde{\xi^2}$ is given in (\ref{barxiSon}) and the collisional term is
\be
I_2^{(C)}=\int d\mathbf{c}_1c_{1x}c_1^2\int d\mathbf{c}_2
\bar{T}_0(\mathbf{c}_1,\mathbf{c}_2)(1+P_{12})\chi_s(c_1)\chi_s(c_2)c_{1x}. 
\ee
Taking into account 
\begin{eqnarray}
&&(b_{\sigma}-1)(c_1^2c_{1x}+c_2^2c_{2x})=\left(\frac{1+\alpha}{2}\right)^2
(\hat{\boldsymbol{\sigma}}\cdot\mathbf{c}_{12})^2[c_{1x}+c_{2x}
+2(\hat{\boldsymbol{\sigma}}\cdot\mathbf{c}_{1})\hat{\sigma}_x
+2(\hat{\boldsymbol{\sigma}}\cdot\mathbf{c}_{2})\hat{\sigma}_x]\nonumber\\
&&-\frac{1+\alpha}{2}(\hat{\boldsymbol{\sigma}}\cdot\mathbf{c}_{12})
[c_1^2\hat{\sigma}_x-c_2^2\hat{\sigma}_x
+2(\hat{\boldsymbol{\sigma}}\cdot\mathbf{c}_{1})c_{1x}
-2(\hat{\boldsymbol{\sigma}}\cdot\mathbf{c}_{2})c_{2x}], 
\end{eqnarray}
and the solid angle integrals
\begin{eqnarray}
\int d\hat{\boldsymbol{\sigma}}
\Theta(\hat{\boldsymbol{\sigma}}\cdot\mathbf{c}_{12})
(\hat{\boldsymbol{\sigma}}\cdot\mathbf{c}_{12})^3
&=&\frac{\pi^{\frac{d-1}{2}}}{\Gamma\left(\frac{d+3}{2}\right)}
c_{12}^3, \\
\int d\hat{\boldsymbol{\sigma}}
\Theta(\hat{\boldsymbol{\sigma}}\cdot\mathbf{c}_{12})
(\hat{\boldsymbol{\sigma}}\cdot\mathbf{c}_{12})^3
(\hat{\boldsymbol{\sigma}}\cdot\mathbf{c}_{1})\hat{\sigma}_x
&=&\frac{\pi^{\frac{d-1}{2}}}{2\Gamma\left(\frac{d+5}{2}\right)}
[c_{12}^3c_{1x}+3c_{12}c_{12x}(\mathbf{c}_1\cdot\mathbf{c}_{12})], \nonumber\\
\end{eqnarray}
we have
\be
I_2^{(C)}
=\frac{(1+\alpha)\pi^{\frac{d-1}{2}}}{2\Gamma\left(\frac{d+3}{2}\right)}
\int d\mathbf{c}_1\int d\mathbf{c}_2\chi_s(c_1)\chi_s(c_2)c_{1x}
\mathcal{F}(\mathbf{c}_1,\mathbf{c}_2), 
\ee
where
\begin{eqnarray}
&&\mathcal{F}(\mathbf{c}_1,\mathbf{c}_2)=\frac{1+\alpha}{d+3}
[c_{12}^3(c_{1x}+c_{2x})+3c_{12}c_{12x}(\mathbf{c}_1+\mathbf{c}_2)
\cdot\mathbf{c}_{12}]-(c_1^2-c_2^2)c_{12}c_{12x}\nonumber\\
&&+\frac{1+\alpha}{2}c_{12}^3(c_{1x}+c_{2x})
+2c_{12}[(\mathbf{c}_{12}\cdot\mathbf{c}_2)c_{2x}
-(\mathbf{c}_{12}\cdot\mathbf{c}_1)c_{1x}]. 
\end{eqnarray}
Evaluating the integral in the first Sonine approximation we get
\be
I_2^{(C)}=-\frac{(1+\alpha)\pi^{\frac{d-1}{2}}}{32\sqrt{2}d\Gamma(d/2)}
\{(32+16d)(1-\alpha)+a_2^s[70+47d-3(34+5d)\alpha]\}. 
\ee
The eigenvalue can be finally written as
\be
\lambda_{NH}^{(2)}=\frac{4}{(d+2)a_2^s}
\left(\frac{d+2}{2}\widetilde{\xi^2}+I_2^{(C)}\right). 
\ee
Let us note that the numerator goes as $a_2^s$ in the elastic limit, so that a 
finite result is obtained in that limit.


\begin{thebibliography}{10}

\bibitem{c90}
C.~S. Campbell, Annu. Rev. Fluid Mech. {\bf 22},  57  (1990).

\bibitem{h83}
P.~K. Haff, J. Fluid. Mech {\bf 134},  401  (1983).

\bibitem{l05}
J.~F. Lutsko, Phys. Rev. E {\bf 72},  021306  (2005).

\bibitem{gs95}
A. Goldshtein and M. Shapiro, J. Fluid. Mech. {\bf 282},  75  (1995).

\bibitem{bds97}
J.~J. Brey, J.~W. Dufty, and A. Santos, J. Stat. Phys. {\bf 87},  1051  (1997).

\bibitem{resibois}
P. R\'esibois and M. de~Leener, {\em Classical Kinetic Theory of Fluids} (John
  Wiley, New York, 1977).

\bibitem{bdks98}
J.~J. Brey, J.~W. Dufty, C.~S. Kim, and A. Santos, Phys. Rev. E {\bf 58},  4638
   (1998).

\bibitem{db03}
J.~W. Dufty and J.~J. Brey, Phys. Rev. E {\bf 68},  030302(R)  (2003).

\bibitem{bdr03}
J.~J. Brey, J.~W. Dufty, and M.~J. Ruiz-Montero,  in {\em Granular Gas
  Dynamics}, edited by T. Poeschel and N. Brilliantov (Springer, Berlin, 2003).

\bibitem{bd05}
J.~J. Brey and J.~W. Dufty, Phys. Rev. E {\bf 72},  011303  (2005).

\bibitem{db02}
J.~W. Dufty and J.~J. Brey, J. Stat. Phys. {\bf 109},  433  (2002).

\bibitem{dab08}
J.~W. Dufty, A. Baskaran, and J.~J. Brey, Phys. Rev. E {\bf 77},  031310
  (2008).

\bibitem{adb08}
A. Baskaran, J.~W. Dufty, and J.~J. Brey, Phys. Rev. E {\bf 77},  031311
  (2008).

\bibitem{l06}
J.~F. Lutsko, Phys. Rev. E {\bf 73},  021302  (2006).

\bibitem{g06}
V. Garz\'o, Phys. Rev. E {\bf 73},  021304  (2006).

\bibitem{peu02}
A. Prevost, D.~A. Egolf, and J.~S. Urbach, Phys. Rev. Lett. {\bf 89},  084301
  (2002).

\bibitem{lvu09}
A.~E. Lobkovsky, F.~V. Reyes, and J.~S. Urbach, Eur. Phys. J. Spec. Top. {\bf
  179},  123  (2009).

\bibitem{pggsv12}
A. Puglisi, A. Gnoli, G. Gradenigo, A. Sarracino, and D. Villamaina, J. Chem.
  Phys {\bf 136},  014704  (2012).

\bibitem{gtsh12}
V. Garz\'o, S. Tenneti, S. Subramaniam, and C. M. Hrenya, J. Fluid Mech. (to be 
published).


\bibitem{clh00}
R. Cafiero, S. Luding, and H.~J. Herrmann, Phys. Rev. Lett. {\bf 84},  6014
  (2000).

\bibitem{wm96}
D.~R.~M. Williams and F.~C. MacKintosh, Phys. Rev. E {\bf 54},  R9  (1996).

\bibitem{vne98}
T.~P.~C. van Noije and M.~H. Ernst, Granular Matter {\bf 1},  57  (1998).

\bibitem{vnetp99}
T.~P.~C. van Noije, M.~H. Ernst, E. Trizac, and I. Pagonabarraga, Phys. Rev. E
  {\bf 59},  4326  (1999).

\bibitem{PTvNE01}
I. Pagonabarraga, E. Trizac, T. P. C. van Noije, and M. H. Ernst,
Phys. Rev. E {\bf 65}, 011303 (2001). 

\bibitem{ms00}
J.~M. Montanero and A. Santos, Granular Matter {\bf 2},  53  (2000).

\bibitem{gm02}
V. Garz{\'o} and J.~M. Montanero, Physica A {\bf 313},  336  (2002).

\bibitem{etb06}  
M. H. Ernst,   E. Trizac and   A. Barrat,  
Europhys. Lett. {\bf 76}, 56 (2006).

\bibitem{vpbtvw06}
P. Visco, A. Puglisi, A. Barrat, E. Trizac, and F. van Wijland, 
Europhys. Lett. {\bf 72}, 55 (2005);
J. Stat. Phys.  {\bf 125},  533  (2006).

\bibitem{vpbvwt06}
P. Visco, A. Puglisi, A. Barrat, F. van Wijland, and E. Trizac, Eur. Phys. J. B
  {\bf 51},  377  (2006).

\bibitem{gmt09}
M.~I. {Garc\'ia de Soria}, P. Maynar, and E. Trizac, Molec. Phys. {\bf 107},
  383  (2009).

\bibitem{mgt09}
P. Maynar, M.~I. {Garc\'ia de Soria}, and E. Trizac, Eur. Phys. J. Special
  Topics {\bf 179},  123  (2009).

\bibitem{vaz11}
K. Vollmayr-Lee, T. Aspelmeier, and A. Zippelius, Phys. Rev. E {\bf 83},
  011301  (2011).

\bibitem{plmv99v}
For a variant of the model, see 
A. Puglisi, V. Loreto, U.~M.~B. Marconi, and A. Vulpiani, Phys. Rev. E {\bf
  59},  5582  (1999).

\bibitem{Bada2012}
V. Garz\'o, M. G. Chamorro and F. V. Reyes, arXiv:1211.4985.
  
\bibitem{gmt12}
M.~I. {Garc\'ia de Soria}, P. Maynar, and E. Trizac, Phys. Rev. E {\bf 85},
  051301  (2012).

\bibitem{vk92}
N.~G. van Kampen, {\em Stochastic Processes in Physics and Chemistry}
  (North-Holland, Amsterdam, 1992).

\bibitem{as07}
A. Astillero and A. Santos, Europhys. Lett. {\bf 78},  1  (2007).

\bibitem{as12}
A. Astillero and A. Santos, Phys. Rev. E {\bf 85},  021302  (2012).

\bibitem{arnoldVasilii}
A.~F. Nikiforov and V.~B. Uvarov, {\em Special Functions of Mathematical
  Physics} (Birkh\"auser Verlag, Basel, 1988).

\bibitem{mg11}
P. Maynar and M.~I. {Garc\'ia de Soria}, Math. Model. Nat. Phenom. {\bf 6},  87
   (2011).

\bibitem{mcLennan}
J.~A. McLennan, {\em Introduction to Nonequilibrium Statistical Mechanics}
  (Prentice-Hall, New Jersey, 1989).

\bibitem{bmg12}
J.~J. Brey, P. Maynar, and M.~I. {Garc\'ia de Soria}, In preparation  .

\bibitem{bmg11}
J.~J. Brey, P. Maynar, and M.~I. {Garc\'ia de Soria}, Phys. Rev. E {\bf 83},
  041303  (2011).

\bibitem{g11}
V. Garz\'o, Phys. Rev. E {\bf 84}, 012301 (2011). 

\bibitem{HansenMcDonald}
J.~P. Hansen and I.~R. McDonald, {\em Theory of Simple Liquids} (Academic
  Press, Amsterdam, 2006).




\end{thebibliography}
\end{document}